\documentclass{article}

\usepackage[a4paper, total={6in, 8in}]{geometry}
\usepackage{authblk}
\usepackage{wrapfig}  
\usepackage{floatflt}
\usepackage[utf8]{inputenc}
\usepackage{amsthm}
\usepackage{amsmath}
\usepackage{amsfonts}
\usepackage{verbatim}
\usepackage{tikz}
\usepackage{multirow}
\usepackage{algorithm2e}
\usepackage{graphicx}
\usepackage{caption}
\usepackage{subcaption}
\usepackage{amssymb}
\usepackage{appendix}
\usepackage{enumerate}
\usepackage{xcolor}
\usepackage{lipsum}
\usepackage{subfiles}
\usepackage{adjustbox}
\usepackage{varwidth}
\usepackage{mdframed}
\usepackage[most]{tcolorbox}
\tcbuselibrary{breakable}

\newcommand{\Tau}{{\mathcal{J}}}

\newcommand{\chord}{{\mathcal{\bf H}}}
\newcommand{\hb}{\hookrightarrow}

\newtheorem{constraint}{Constraint}
\newtheorem{claim}{Claim}
\newtheorem{lemma}{Lemma}
\newtheorem{theorem}{Theorem}
\newtheorem{definition}{Definition}

\AtBeginDocument{%
  \providecommand\BibTeX{{%
    \normalfont B\kern-0.5em{\scshape i\kern-0.25em b}\kern-0.8em\TeX}}}

\begin{document}

\title{Partially Replicated Causally Consistent Shared Memory:\\
 Lower Bounds and An Algorithm 
 \thanks{This research is supported in part by National Science Foundation award 1409416, and Toyota InfoTechnology Center. Any opinions, findings, and conclusions or recommendations expressed here are those of the authors and do not necessarily reflect the views of the funding agencies or the U.S. government.}}


\author[1]{Zhuolun Xiang\thanks{xiangzl@illinois.edu}}
\author[2]{Nitin H. Vaidya\thanks{nitin.vaidya@georgetown.edu}}
\affil[1]{University of Illinois at Urbana-Champaign}
\affil[2]{Georgetown University}

\date{}

\maketitle

\begin{abstract}
	The focus of this paper is on causal consistency in a {\em partially
	replicated} distributed shared memory (DSM) system that provides the abstraction
	of shared read/write registers. 
	Maintaining causal consistency in distributed shared memory systems has received 
	significant attention in the past, mostly
	on {\em full replication} wherein each replica stores a
	copy of all the registers in the shared memory.
	To ensure causal consistency, all causally preceding updates must 
	be performed before an update is performed at any given replica.
	Therefore, some mechanism for tracking causal dependencies is required, such as vector timestamps
	with the number of vector elements being equal to the number of replicas in the context of full replication.
	In this paper, we investigate causal consistency in {\em partially replicated systems}, wherein
	each replica may store only a subset of the shared registers.
	Building on the past work,
	this paper makes three key contributions:
	\begin{itemize}
		\item We present a necessary condition on the metadata (which
		we refer as a {\em timestamp})
		that must be maintained by each replica to be able to track causality accurately.
		The necessary condition identifies a set of directed edges in a {\em share graph} that a replica's timestamp must keep track of.
		
		\item We present an algorithm for achieving causal consistency using
		a timestamp that matches the above necessary condition,
		thus showing that the condition is necessary and sufficient.
		
		\item We define a measurement of timestamp space size and present a lower bound (in bits) on the size of the timestamps. The lower bound matches our algorithm in several special cases.
	\end{itemize}
\end{abstract}



	\section{Introduction}
	\label{sec:intro}
\begin{figure}[htp]
	\captionsetup[subfigure]{justification=centering}
	\centering
	\begin{subfigure}[b]{0.4\textwidth}
		\includegraphics[width=\textwidth]{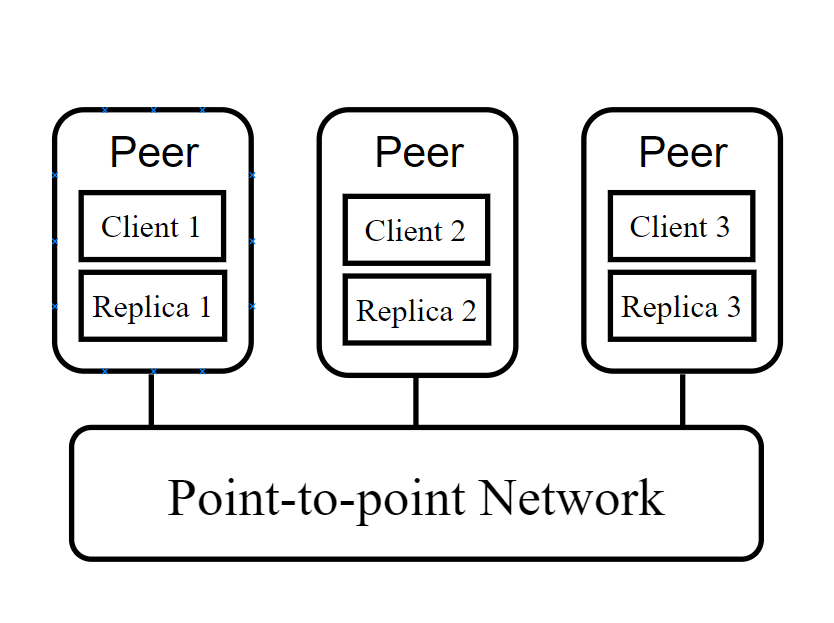}
		\caption{Peer-to-peer architecture}
		\label{fig:single}
	\end{subfigure}
	\begin{subfigure}[b]{0.4\textwidth}
		\includegraphics[width=\textwidth]{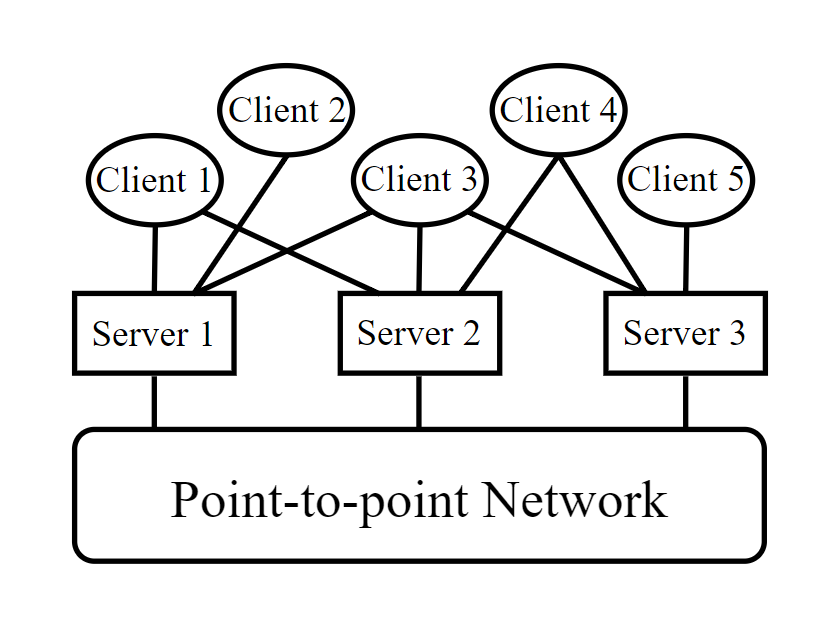}
		\caption{Client-server architecture}
		\label{fig:multiple}
	\end{subfigure}
	\label{fig:dsm}
\end{figure}

	Distributed shared memory systems maintain multiple replicas of the shared memory locations, which we refer as shared registers.
	In recent years, the {\em causal consistency} model for the shared memory has received significant attention due to its emerging applications, such as social networking.
	Intuitively, causal consistency ensures that before an update is applied to a shared register, all the causally preceding updates must be applied at the same replica.
	This paper mainly focuses on the architecture illustrated in Figure \ref{fig:single}, which we refer as the {\em peer-to-peer} architecture.
Each peer has a {\em client} that issues read/write operations to the shared memory and a {\em replica} that helps implement the shared memory abstraction. We focus on the case
when each replica is {\em partial} and may store a copy of just a subset of the shared registers. {\em Full replication} is obtained as a special case when
each replica stores a copy of each shared register.

	We primarily present the results for
	the peer-to-peer architecture. The results easily extend
	to the client-server architecture  in Figure \ref{fig:multiple} where each client may be accessing replicas stored at an arbitrary subset of the servers, as briefly
	discussed in Section \ref{sec:general}.

	In the context of {\em full replication},
	several causally consistent shared memory systems have been
	designed, including
	Lazy Replication \cite{Ladin1992ProvidingHA}, COPS \cite{Lloyd2011DontSF}, Orbe \cite{Du2013OrbeSC}, SwiftCloud \cite{zawirski2015write} and GentleRain \cite{Du2014GentleRainCA}.
	Recently, there is also growing interest in {\em partial replication}
	due to the potential storage
	efficiencies that can be attained
 \cite{bravo2015towards, mahmood2016achieving,Bailis2012ThePD,Hlary2006AboutTE,crain2015designing, dahlin2006practi,  Kshemkalyani2015ApproximateCC, hsu2016performance, mehdi2017can, bravo2017saturn}.
	For {\em full replication}, 
	it suffices to use
	a vector timestamp \cite{Mattern1988VirtualTA,fidge1987timestamps,charron1991concerning} of length equal to the number of replicas
	\cite{Ladin1992ProvidingHA} to achieve causal consistency.

	Several researchers have observed that {\em partial replication} requires larger amount of metadata to track causal dependencies \cite{Bailis2012ThePD,Lloyd2011DontSF,Hlary2006AboutTE,crain2015designing}. 
  For partial replication, in general, the timestamp (or metadata) overhead is expected to be larger than that for full replication in order to avoid {\em false dependencies} as will be explained below.
  One straightforward method to implement partial replication is by adding ``virtual registers'' at each replica to simulate full replication.
  The virtual registers do not store actual data and cannot be accessed by clients.
  Then solutions for full replication such as vector clocks can be easily adapted for partial replication. However, there are several issues: (1) Every update message with metadata will be sent to all replicas in full replication, which is not necessary for partial replication. This may result in high bandwidth usage. (2) Simulating full replication introduces unnecessary dependencies (which we call {\em false dependencies}) among the update messages. For instance, if update $u_x$ on register $x$ depends on update $u_y$ on register $y$, i.e. $u_x$ can only be applied after $u_y$ is applied, then on any replica who received $u_x$ first will wait for the receipt of $u_y$, even if register $y$ is virtual and not stored locally. However, there is no reason for such delay, since the virtual register $y$ will not be accessed by any client from this replica, and thus $u_x$ can be applied without the receipt of $u_y$. 
  Therefore this simulation approach may result in stale versions.

	Partial replication yields a trade-off between flexibility of replication,
	number of false dependencies on update messages,
	and overhead of the metadata for tracking causality.
	A goal of our work is to characterize this trade-off. 
	Intuitively, in our solution, each replica maintains an {\em edge-indexed vector timestamp} which keeps counters for a subset of edges in a {\em ``share graph''} that characterizes how registers are shared among the replicas. We show that our timestamp is optimal in the sense that the subset of share graph edges tracked is necessary for correctness (Theorem \ref{thm:Ei}). 
	Also there is no false dependency introduced in our solution.
	Our main contributions are as follows:

\begin{itemize}
\item We present a necessary condition on the metadata
that must be maintained by each replica to be able to track causality accurately.
The necessary condition identifies a set of directed edges in a share graph that a replica's timestamp must keep track of.

In deriving the necessary condition, we make improvements over results presented in prior work of H{\'e}lary and Milani \cite{Hlary2006AboutTE,Xphdthesis}.

\item We present an algorithm for achieving causal consistency using
a timestamp that matches the above necessary condition,
thus showing that the condition is necessary and sufficient.

\item We define a measurement of timstamp space size and present a lower bound (in bits) on the size of the timestamps. The lower bound matches our algorithm in several special cases.

        \end{itemize}

	\section{Preliminaries}\label{sec:pre}
	
	We assume an asynchronous system, and the replicas communicate using reliable point-to-point message-passing channels.
	The communication channels are not necessarily FIFO.
	In Sections \ref{sec:pre} through \ref{sec:bounds}, we assume the peer-to-peer architecture in Figure \ref{fig:single}.
	Each peer contains a client and a replica. 
	There are $R$ peers, and hence there are $R$ replicas. The
	replicas are numbered 1 through $R$.
	Replica $i$ stores copies of a subset of shared registers named $X_i$.  
	With full replication, $X_i=X_j$ for all replicas $i,j$.
	With partial replication, it is possible that $X_i\neq X_j$ for $i\neq j$. 
	We define $X_{ij}=X_i\cap X_j$, the set of registers stored at replicas $i$ and $j$ both. 
	For instance,in partial replication with four replicas, we may have $X_1=\{x\}$, $X_2=\{x,y\}$, $X_3=\{y,z\}$, and $X_4=\{z\}$, where $x,y,z$ are registers. In this case, $X_{23}=\{y\}$ and $X_{14}=\emptyset$.
	In practice, 
	set $X_r$ for replica $r$ may change dynamically, however, we consider the {\em static} case in this paper and leave the {\em dynamic} case for future work.


H{\'e}lary and Milani \cite{Hlary2006AboutTE} introduced
the notion of a {\em share graph} to represent a partially
replicated system. Similar notions of {\em graph of groups} are introduced in causal multicast literature as well \cite{birman1991lightweight}. We will use the share graph when obtaining results for the {\em peer-to-peer} architecture in Section \ref{sec:conditions} and \ref{sec:bounds}. To extend these results to the {\em client-server} architecture, in Section \ref{sec:general}, we will introduce an {\em augmented} version of the share graph.

\begin{definition}[\bf Share Graph \cite{Hlary2006AboutTE}]\label{def:sg}
	We denote $e_{ij}$ as a directed edge from $i$ to $j$.
	Share graph is defined as a directed graph $G=(V,E)$, where $V=\{1,2,\cdots,R\}$, and
	vertex $i\in V$ represents replica $i$. There exist directed edges $e_{ij}$ and $e_{ji}$
	in $E$ if and only if $X_{ij}\neq \emptyset$.
\end{definition}

As such, if $e_{ij}\in E$ then $e_{ji}\in E$, and $G$ may be defined as an 	undirected graph as originally defined in  \cite{Hlary2006AboutTE}. However, as seen later, it is convenient to represent the sharing using pairs of directed edges.	

$X_{ij}$ will be referred as the label of edges $e_{ij}$ and $e_{ji}$.
In this paper, we assume that {\em each replica has the knowledge of the share graph including the labels on on each edge}.
	
	\begin{wrapfigure}[6]{r}{0.4\textwidth}
	\vspace{-\baselineskip}
		\includegraphics[width=0.4\textwidth]{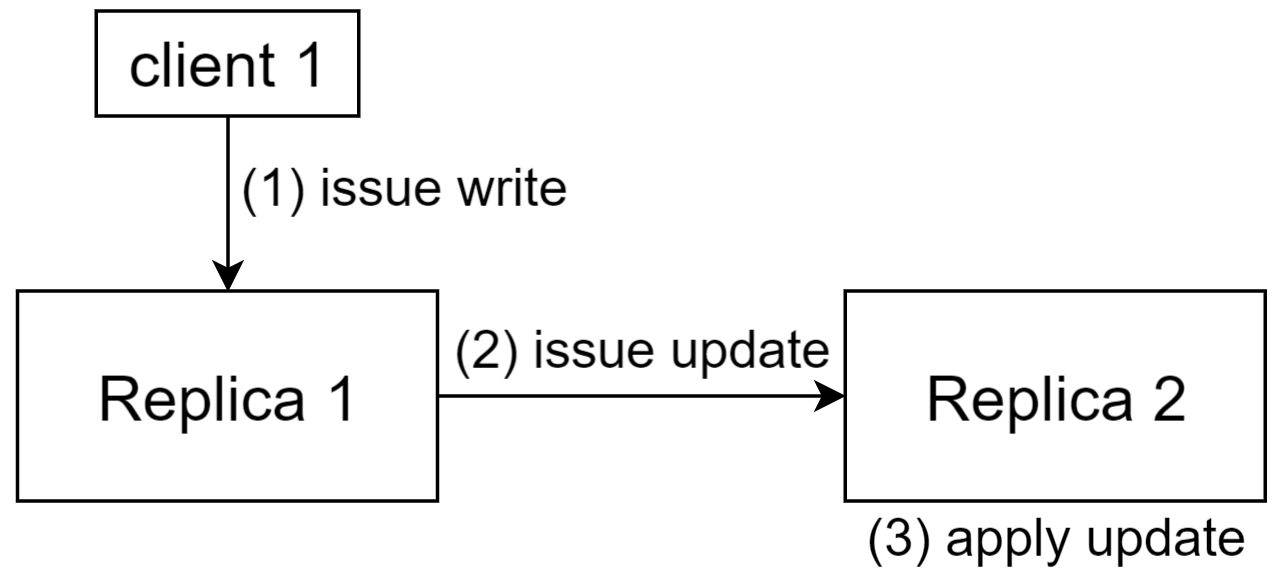}
		\caption{Illustration}
		\label{fig:apply}
	\end{wrapfigure}

	Recall that each peer contains a client and a replica, and
	the client can issue {\em read} or {\em write} operations on a shared register stored at the local replica.
	Define $read(x)$ to be a read operation on register $x$, and $write(x, v)$ to be a write
	operation on register $x$ that writes value $v$. When performing $read(x)$ or $write(x, v)$ operation on register $x\in X_i$
	, client $i$ sends a request to a replica $i$, and awaits the replica’s
	response. The response to a write operation is an acknowledgement, and the response to a
	read operation is a returned value.
Define update to be a tuple of the form $update(i,T,x,v)$, where $i$ is the sender of the update, $T$ is the timestamp attached with the update, $x$ is the register being updated and $v$ is the value.
	As illustrated in Figure \ref{fig:apply},
	upon receiving {\em write} operations from the client, the replica will {\em issue updates} to some other replicas, i.e., sending the tuple $update(i,T,x,v)$ to other replica who also replicates $x$ in order to update their registers. 
	Upon receiving update tuple $update(i,T,x,v)$ from other replica, the replica can decide when to {\em apply the update}, i.e., write the new value $v$ into the register $x$.
	An {\em execution} is defined to be a sequence of clients' read/write operations and replicas' operations in issuing/applying updates. With our definition of replica-centric causal consistency in the next section, we will often construct executions by declaring the replicas' operations on updates without explicitly mentioning the clients' operations.
	

	\subsection{Replica-centric Causal Consistency}\label{sec:causal}
	In this section, we will use the notions of
	{\em issuing an update} and {\em applying an update} mentioned above.
	A client $i$ may only read/write registers in $X_i$. Thus, replica $i$ may only issue updates to registers in $X_i$.
	For convenience, each {\em write} operation on a given register is assumed to write a unique value. 
	
	In past work, several variations of causal consistency have been explored. One of the commonly used definition of causal consistency is defined from clients' view point, which we refer as {\em client-centric causal consistency} below.
	
	\begin{definition}[Happened-before relation $\rightarrow$ for operations \cite{lamport1978time}]
		Let $o_1,o_2$ be two operations of the client. $o_1$ happened-before $o_2$, denoted as $o_1\rightarrow o_2$, if and only if at least one the following conditions is true:
		(1) Both $o_1$ and $o_2$ are performed by the same client, and $o_1$ occurs before $o_2$.
		(2) $o_1$ is a write operation, and $o_2$ is a read operation that returns the value written by $o_1$.
		(3) There exists operation $o_3$ such that $o_1\rightarrow o_3$ and $o_3\rightarrow o_2$.
	\end{definition}
	Client-centric causal consistency is defined based on the relation $\rightarrow$ for operations above.
	
	\begin{definition}[Client-centric Causal Consistency]
		\label{def:causal}
		Client-centric causal consistency is achieved if the following two properties are satisfied:
		\begin{itemize}
			\item {\bf Safety:} If a read operation $o_3$ on some register $x$ returns the value written
			by write operation $o_1$ on register $x$, then there {\em must not exist} another
			write operation $o_2$ on register $x$ such that $o_1\rightarrow o_2\rightarrow o_3$.
			\item {\bf Liveness:} For a write operation $o_1$ by some client that writes
			value $v$ in register $x$, all replicas
			that store copies of register $x$ should be updated with the value $v$ within a finite time.
		\end{itemize}
	\end{definition}

	The causal consistency model addressed in our work is inspired by replicated shared memory systems such as
	Lazy Replication \cite{Ladin1992ProvidingHA}. We refer to this model
	as the {\em replica-centric} causal consistency model.  We define the {\em happened-before} relation \cite{lamport1978time} between updates as follows.
	
	\begin{definition}[Happened-before relation $\hb$ for updates]
		\label{def:hb}
		Given updates $u_1$ and $u_2$, $u_1\hb u_2$ if and only if at least one of the following conditions is true: 
		\begin{enumerate}
			\item $u_1$ is applied at a replica on any of its register sometime before the same replica issues $u_2$ on any of its register. 
			\item There exists an update $u_3$ such that $u_1\hb u_3$ and $u_3\hb u_2$.
		\end{enumerate}
		
	\end{definition}

	\begin{wrapfigure}{r}{0.4\textwidth}
	\vspace{-\baselineskip}
		\includegraphics[width=0.4\textwidth]{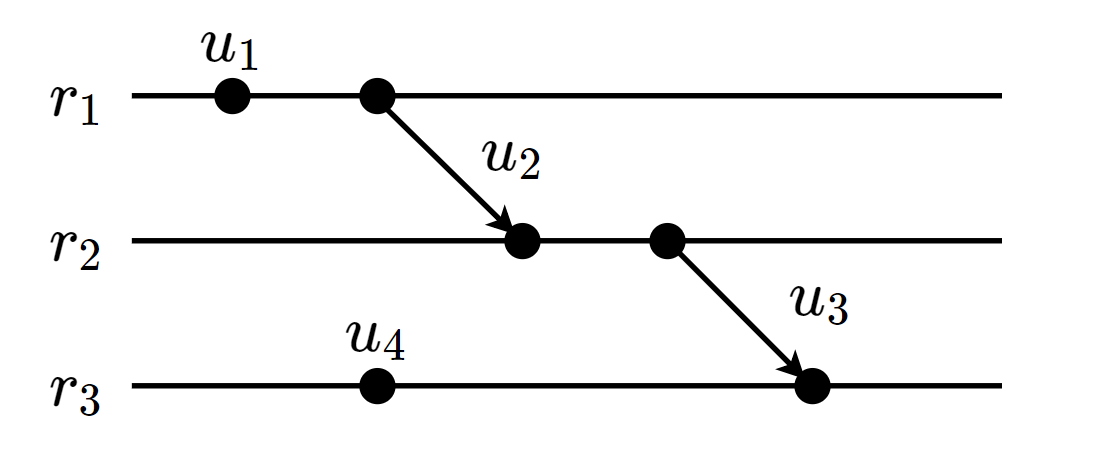}
		\caption{Relation $\hb$}
		\label{fig:hb}
	\end{wrapfigure}
	
	Intuitively, relation $\hb$  is analogous to the happened-before relation between events in the context of causal multicast. 
	That is, an update issued by replica $i$ is considered causally dependent
	on any updates that were previously applied at that replica, regardless of whether the
	previously updated registers were read by the client or not.

	We give an example of relation $\hb$ in Figure \ref{fig:hb}.
	In this example, there are $3$ replicas $r_1,r_2,r_3$, where $r_1$ issues updates $u_1$ and $u_2$, $r_2$ issues update $u_3$ and $r_3$ issues update $u_4$. $u_1$ is applied at $r_1$, $u_2$ is applied at $r_1,r_2$, $u_3$ is applied at $r_2,r_3$ and $u_4$ is applied at $r_3$.
	In the figure, the send event of arrow with label $u_2$ depicts when update $u_2$ is issued at $r_1$, and the receive event of that arrow depicts the time when $r_2$ applies update $u_2$.
	By condition (1) of the $\hb$ definition, we have $u_1\hb u_2$ and $u_2\hb u_3$, and
	by condition (2) we have $u_1\hb u_3$. Also, $u_1$ and $u_4$ are concurrent, i.e. $u_1\not\hb u_4$ and $u_4\not\hb u_1$. Similarly, $u_2$ and $u_4$ are concurrent.

	We define {\em replica-centric}
	causal consistency formally now using relation $\hb$.
	
	\begin{definition}[Replica-centric Causal consistency]
		\label{def:causal2}
		Replica-centric causal consistency is achieved if the following two properties are satisfied:
		\begin{itemize}
			\setlength{\itemsep}{0pt}
			\item {\bf Safety:} If an update $u_1$ for register $x\in X_i$ has been applied at a
			replica $i$, then there {\em must not exist} update $u_2$ for some register in $X_i$ such that (i)  $u_2\hb u_1$, and (ii) replica $i$ has not yet applied $u_2$.
			
			\item {\bf Liveness:\footnote{Note that our definition of Liveness implies no false dependencies.}} Any update $u$ issued by a replica $i$ for a register $x\in X_i$ should be applied at each replica $j$ such that $x\in X_j$ within a finite time after all dependencies of $u$ have been applied at $j$, i.e., all $u'$ for some register $y\in X_j$ such that $u'\hb u$ have been applied.
		\end{itemize}
		
	\end{definition}
	
	For three reasons, we consider the replica-centric causal consistency in this paper.  
	(i) First, the necessary conditions presented in Section \ref{sec:nec} and \ref{sec:bounds} for the replica-centric causal consistency also applies to the client-centric causal consistency.
	(ii) Second, the algorithms for replica-centric causal consistency also implement the client-centric causal consistency, but with possible false dependencies.
	(iii) Third, in practice, maintaining the replica-centric causal consistency is efficient in metadata size, since it only uses a single timestamp per replica (as compared to, for instance, a timestamp per register per replica for the client-centric causal consistency). Many practical systems, including Lazy Replication \cite{Ladin1992ProvidingHA}, ChainReaction \cite{almeida2013chainreaction} and SwiftCloud \cite{zawirski2015write}, in fact, conform to the replica-centric view.


	\paragraph*{Relation with Causal Group Multicast} As we mentioned earlier,  the definition of replica-centric causal consistency is analogous to the requirement for causal group multicast \cite{birman1991lightweight}, where the messages need to be delivered to the processes in a causal order.  
	The following correspondence can be obtained. Replicas sharing the same register $x$ correspond to processes belonging to the same multicast group $G_x$. Any update to register $x$ by replica $i$ results in a multicast to group $G_x$ by replica $i$.
	In the case of partial replication, our algorithm later in section \ref{sec:suf} can essentially be viewed as  causal group multicast with {\em overlapping groups} \cite{birman1991lightweight, mostefaoui1993causal, kshemkalyani1998necessary}, where each process may belong to multiple groups (determined by how they share registers) and the multicast within a group is only received by members in that group. 
	Hence our results below in Section \ref{sec:nec} and \ref{sec:suf} also apply to causal multicast with overlapping groups. For the sake of the consistency of presentation, we state our results in the context of distributed shared memory. Related work on causal group multicast and the comparison with our work are discussed in Section \ref{sec:relatedwork}.
	
	\begin{tcolorbox}[breakable, enhanced]
		In this paper, we consider algorithms that implement causal consistency by storing and attaching metadata (timestamps) with update messages, where the metadata is some encoding of the information about the execution history. When to apply a received update at a replica is determined only using the replica's local metadata and the metadata attached with the update.
	\end{tcolorbox}

\section{Timestamps for Replica-Centric Causal Consistency}
\label{sec:conditions}

In this section, we consider partially replicated shared memory systems, which
satisfy the replica-centric causal consistency model in Section \ref{sec:causal}
using an algorithm under the assumptions mentioned in Section \ref{sec:pre}.
In particular, we identify a necessary and sufficient condition on the
timestamp $\tau_i$ maintained by each replica $i$.
Intuitively, our condition identifies a subset of directed edges in the share graph that are {\em necessary and sufficient} to ``keep track'' of for each replica in order to achieve replica-centric causal consistency.

\begin{wrapfigure}{r}{0.3\textwidth}
\vspace{-\baselineskip}
	\includegraphics[width=0.3\textwidth]{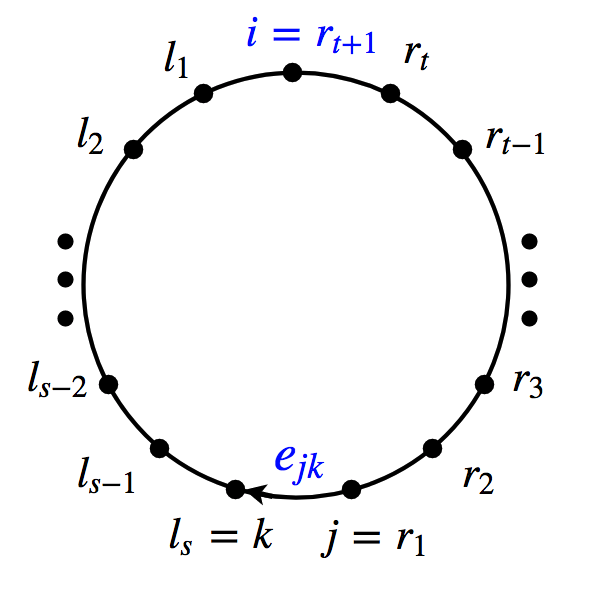}
	\caption{$(i,e_{jk})$-loop}
	\label{fig:ieloop}
\end{wrapfigure}

	For a replica $i$, and directed edge $e_{jk}$ (from $j$ to $k$) in the share graph, Definition \ref{def:ijk} defines an $(i,e_{jk})$-loop as illustrated in Figure \ref{fig:ieloop}.
	We will use $-$ to denote the set difference, i.e., $A-B=\{x\in A~|~x\notin B\}$.
	After introducing the definition below, we provide an intuition behind the definition.
	
	\begin{definition}[\bf $(i,e_{jk})$-loop]
		\label{def:ijk}
		Given replica $i$ and edge $e_{jk}$ ($j\neq i\neq k$) in share graph $G$,
		consider a simple loop of the form
		$(i,\,l_1,\,l_2, \cdots,l_s=k,\,j=r_1,\, r_2,\cdots,r_t,\,i)$, where
		$s\geq 1$ and $t\geq 1$.
		Define $i=r_{t+1}$.
		The simple loop is said to be an $(i,e_{jk})$-loop provided that:\\
		\indent (i) $X_{jk}  -  \left( \cup_{1 \leq p\leq s-1} \,X_{l_p}\right)\neq\emptyset$, \\
		\indent (ii) $X_{jr_2}  -  \left( \cup_{1 \leq p\leq s-1} \,X_{l_p}\right)\neq\emptyset$, and\\
		\indent (iii) for $2\leq q\leq t$,  $X_{r_q r_{q+1}}  -  \left( \cup_{1\leq p\leq s}\, X_{l_p}\right)\neq\emptyset$.
	\end{definition}

As shown later, when there exists an $(i,e_{jk})$-loop, replica $i$ need to keeps information regarding updates on edge $e_{jk}$ in order to achieve causal consistency.

	{\bf Intuition:}
	This discussion refers to Figure \ref{def:ijk}.
	The definition of $(i,e_{jk})$-loop will be used to characterize the timestamp used by replica $i$ in our algorithm. 
	As we will show later in the proof of Theorem \ref{thm:Ei}, the timestamp at replica $i$ must reflect information regarding updates on edge $e_{jk}$ if an $(i,e_{jk})$-loop exists.
	Consider the following execution.
	Let $u$ be an update issued by replica $j$ which is sent to replica $k$ (i.e., an update on edge $e_{jk}$)  since replica $k$ stores the register that $u$ is updating.
	Also suppose that there is a sequence of causally dependent updates propagated along the path $(j, r_2, \cdots, i,\cdots, l_{s-1}, k)$. Denote the update from $l_{s-1}$ to $k$ as $u'$, so that we have $u\hb u'$. Then the timestamps attached with $u'$ should contain enough information about the $u\hb u'$ relation for replica $k$ to apply these two updates in the correct order, or postpone the application of $u'$ if $u'$ is received before $u$.
	Thus, it is necessary for replicas such as replica $i$ to ``keep track of'' causally preceding updates that have taken place on edge $e_{jk}$. This allows replica $i$ to propagate the dependency information to other replicas in the above loop that need it (particularly, replica $l_1$ to $l_{s}$).
	If condition (i) is not true, update $u$ will also be sent to some replica $l_p$ where $1\leq p\leq s-1$, since all registers shared by $j,k$ are also shared by $j$ and some replica $l_p$. 
	Similarly, if condition (ii) or (iii) is not true, the updates along the path $(j,r_2,...,i)$ will also be sent to some replica $l_p$. Since the timestamps of these updates sent to some $l_p$ may contain the information about $u\hb u'$, it is not necessary for replica $i$ to ``keep track of'' the causality for updates on edge $e_{jk}$. On the other hand, when all three conditions are true, replica $i$ has to maintain such information to ensure causal consistency.
	For more details, the reader may refer to the proofs for the necessary and sufficient condition in later sections.

\begin{wrapfigure}{r}{0.41\textwidth}
		\vspace{-1\baselineskip}
		\captionsetup[subfigure]{justification=centering}
		\centering
		\begin{subfigure}[b]{0.2\textwidth}
			\includegraphics[width=\textwidth]{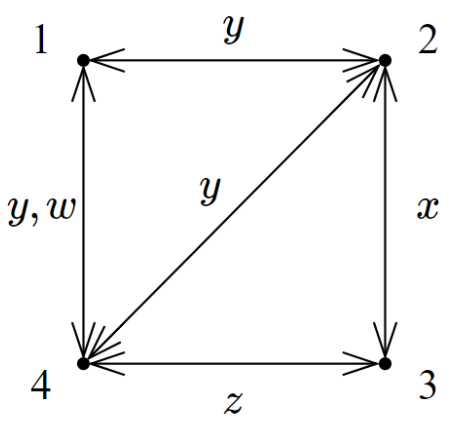}
			\caption{$G$}
			\label{fig:sgex}
		\end{subfigure}
		\begin{subfigure}[b]{0.2\textwidth}
			\includegraphics[width=\textwidth]{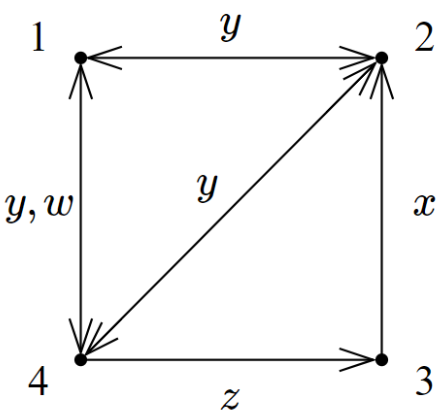}
			\caption{$G_1$}
			\label{fig:GA}
		\end{subfigure}
		\caption{}
		\label{fig:share}
	\end{wrapfigure}

	{\bf Example:} Figure \ref{fig:sgex} shows a share graph for a system of 4 replicas.
	Suppose that $X_1=\{a,y,w\}$, $X_2=\{b,x,y\}$, $X_3=\{c,x,z\}$ and $X_4=\{d,y,z,w\}$.
	The label on edges between replicas $i$ and $j$ in Figure \ref{fig:sgex} corresponds to
	the registers in $X_{ij}$. For instance, $X_{34}=\{z\}$.
	By Definition \ref{def:ijk}, $(1,4,3,2)$ is {\em not} a $(1, e_{34})$-loop since $X_{21}-X_{4}=\emptyset$ which violates condition (iii). 
	Similarly, $(1,4,3,2)$ is {\em not} a $(1, e_{23})$-loop due to a similar reason. On the other hand, $(1,2,3,4)$ is a $(1,e_{43})$-loop. Due to the existence of register $w$ in $X_{14}$, $X_{14}-X_{2}\neq \emptyset$, and the reader can easily check that all three conditions in Definition \ref{def:ijk} are satisfied. Similarly, $(1,2,3,4)$ is a $(1,e_{32})$-loop.

To help present
the necessary condition in Section \ref{sec:nec}, we now define a {\em timestamp graph}.
Intuitively, timestamp graph $G_i$ consists of directed edges that are {\em necessary and sufficient} for replica $i$  to keep track of in its timestamp, as we will show later in Section \ref{sec:nec} and \ref{sec:suf}.

	\begin{definition}[\bf Timestamp graph $G_i$ of replica $i$]
		\label{def:timestamp}
		Given share graph $G=(V, E)$,
		timestamp graph of replica $i$ is defined as a directed graph $G_i=(V_i,E_i)$, where
		\begin{equation*}
		    \begin{aligned}
		    E_i= & \{e_{ij}~|~e_{ij}\in E\}~\cup~ \{e_{ji}~|~e_{ji}\in E\} \\
		    & \cup~ \{e_{jk} ~|~
		~\exists~(i,e_{jk})\text{-loop in~} G,~ j\neq i\neq k,~e_{jk}\in E
		\}
		\\
		V_i=&\{ u,v ~|~ e_{uv}\in E_i \}
		    \end{aligned}
		\end{equation*}
	\end{definition}
	
	Thus, $E_i$ consists of all directed edges incident at $i$, and
	each edge $e_{jk}\in E$ such that an $(i,e_{jk})$-loop exists.
Consider the share graph example in Figure \ref{fig:share}(a) again.
Figure \ref{fig:share}(b) shows the {\em timestamp graph} for replica 1.
Observe that the edge $e_{43}$ is in $G_1$ but $e_{34}$ is not in $G_1$, due to the fact that $(1,2,3,4)$ is a $(1,e_{43})$-loop but $(1,4,3,2)$ is {\em not} a $(1, e_{34})$-loop, as we explained earlier for the example of $(i,e_{jk})$-loop.
By the example above and the definition of timestamp graph, we make the following three observations: 

{
\em 1. Timestamp graphs may be different than the share graph.
2. Different replicas may have different timestamp graphs.
3. Edges in the timestamp graph are not necessarily bidirectional.
}

\subsection{A Necessary Condition for Timestamps}\label{sec:nec}

As briefly stated in Section \ref{sec:pre},
each replica maintains a timestamp. To achieve replica-centric causal consistency,
the timestamp must contain enough information.
In this section, we obtain a necessary condition on the timestamps. In particular, 
Theorem \ref{thm:Ei} below shows that, if $e_{jk}$ is
in the timestamp graph of replica $i$, then it is necessary for replica $i$ to
``keep track of'' updates performed by replica $j$ to registers in $X_{jk}$.
%
%
To present the result formally, we introduce some additional terminology.


	


\begin{definition}[{\bf Causal past and Causal dependency graph \cite{giovannathesis}}] \label{def:cdg}
Causal dependency graph $\mathcal{R}$ of a replica
that has applied updates in set $U$ consists of vertices in
$
{S} = U \cup \{ u'~|~u\in U,~u'\hb u \}
$
and directed edges in
$
\{ (u_1,u_2)~|~u_1\hb u_2\mbox{~and~}u_1,u_2\in S\}
$.
Set ${S}$ is referred as the {\em causal past} of the replica \cite{giovannathesis}.
\end{definition}

Theorem \ref{thm:Ei} will use the following terminology:
\begin{itemize}
	\setlength{\itemsep}{0pt}
	\item We define relation for causal dependency graph $\mathcal{R}_0, \mathcal{R}_1$ of replica $i$ as follows:
	$\mathcal{R}_0 < \mathcal{R}_1$ if there exists an execution in which replica $i$'s causal dependency graph equals $\mathcal{R}_0$ at some point of time, and $\mathcal{R}_1$ subsequently. 
	
	
	\item Two causal dependency graphs with vertex sets $S_1$ and $S_2$ 
	are said to differ only in updates on $e_{jk}$
	if and only if (i) all the updates in $(S_1-S_2)\cup (S_2-S_1)$
	are issued by replica $j$ for registers in $X_{jk}$ and (ii) 
	the edges between any vertices in $S_1\cap S_2$ are identical
	in both the causal dependency graphs.
	
	\item We will say that replica $i$ with causal dependency graph
	$\mathcal{R}_0$ is {\em oblivious} to updates
	on $e_{jk}$, if the replica's timestamp is identical for 
	every pair of causal dependency graphs $\mathcal{R}_1$ and
	$\mathcal{R}_2$ such that (i) $\mathcal{R}_0<\mathcal{R}_1$
	and $\mathcal{R}_0<\mathcal{R}_2$, and (ii) $\mathcal{R}_1$ and $\mathcal{R}_2$
	differ only in updates to $X_{jk}$.
	
\end{itemize}

Intuitively, a replica that is oblivious to updates on $e_{jk}$ does not keep
track of updates to registers in $X_{jk}$ by replica $j$.
\begin{theorem}\label{thm:Ei}
Consider a partially replicated shared memory system that implements replica-centric causal consistency.
Any replica $i$ must not be oblivious to update on any edge $e\in E_i$, where $E_i$ is the edge set in the timestamp graph of replica $i$.
\end{theorem}

The proof of Theorem \ref{thm:Ei} is provided in Appendix \ref{sec:necproof}.
Intuitively, the theorem states that, replica $i$'s timestamp needs to be
dependent on the updates performed on edge $e_{jk}$ for each $e_{jk}\in E_i$.
For instance, a vector timestamp whose elements are indexed by
edges in $E_i$, and count updates performed on the corresponding edges,
satisfies the requirements in Theorem \ref{thm:Ei}. Indeed, in
Section \ref{sec:suf} we present an algorithm that uses precisely such a timestamp,
proving that the necessary condition in Theorem \ref{thm:Ei} is sufficient
as well.
Later in Section \ref{sec:bounds} we obtain a lower bound on the size of the timestamps
in the unit of bits. The necessary condition of Theorem \ref{thm:Ei} does not
provide a measure of the size of the timestamps, whereas Theorem \ref{thm:prlowerbound} provides
lower bound on the size.

\subsection{Sufficiency of Tracking Edges in Timestamp Graph}\label{sec:suf}

We propose an algorithm for implementing causally consistent shared memory in this section.  
The algorithm is for peer-to-peer architecture where each client only issues operations to one corresponding replica.
Recall that $G_i=(V_i,E_i)$ is the timestamp graph of replica $i$.

{\bf Timestamps:}
Each replica $i$ maintains an {\em edge-indexed} vector timestamp $\tau_i$ that
is indexed by the edges in $E_i$. For edge $e_{jk}\in E_i$,
$\tau_i[e_{jk}]$ is an integer, initialized to $0$. 

\begin{tcolorbox}[breakable, enhanced]
	\textbf{Client's Algorithm:}
	\begin{enumerate}
		\item Upon {\em read} operation on register $x$: send $read(x)$ to the replica, wait for the value returned by the replica.
		\item Upon {\em write} operation on register $x$ with value $v$: send $write(x,v)$ to the replica, wait for the acknowledgement from the replica.
	\end{enumerate}
	\textbf{Replica's Algorithm:}
	\begin{enumerate}
		\setlength{\itemsep}{0pt}
		\item Upon receiving a $read(x)$ request from the client:
		
		replica $i$ responds with the value of the local copy of register $x$.
		
		\item Upon receiving a $write(x,v)$ request from the client:
		
		replica $i$ performs the following operations atomically:
		\begin{enumerate}
			\setlength{\itemsep}{0pt}
			\item write $v$ into the local copy of register $x$,
			\item for each $e_{jk}\in E_i$, update timestamp $\tau_i$ as
			
			$\tau_i[e_{jk}]:= \left\{ 
			\begin{array}{l}
			\tau_i[e_{jk}]+1, \mbox{~if~} j=i~\mbox{and~} x\in X_{ik},\\
			\tau_i[e_{jk}], \text{~~~~~ otherwise}
			\end{array}
			\right.
			$
			\item send $update(i,\tau_i,x,v)$ message
			to each other replica $k\in V$ such that $x\in X_k$, 
			\item return {\em ack} to the client.
		\end{enumerate}


		
		\item Upon receiving a message $update(k,\tau_k,x,v)$ from replica $k$:
		
		replica $i$ adds $update(k,\tau_k,x,v)$ to a local data structure named $pending_i$.

		\item
		For any $update(k,\tau_k,x,v)\in pending_i$,
		when $\tau_i[e_{ki}]= \tau_k[e_{ki}]-1$ and 
		$\tau_i[e_{ji}] \geq \tau_k[e_{ji}]$ for each $e_{ji}\in E_i\cap E_k,~i\neq j\neq k$,
		replica $i$ performs the following
		operations atomically: 
		
		\begin{enumerate}
			\setlength{\itemsep}{0pt}
			\item writes value
			$v$ to its local copy of register $x$,
			\item for each $e\in E_i$, updates timestamp $\tau_i$ as
			
			$\tau_i[e]:= \left\{ 
			\begin{array}{l}
			\max \left(\tau_i[e], \tau_k[e]\right),
			\text{~ for each ~} e\in E_i\cap E_k, \\
			\tau_i[e],~~~~~~~~~~~~~~~\, \text{~ for each edge~} e\in E_i-E_k
			\end{array}
			\right.
			$
			\item removes $update(k,\tau_k,x,v)$ from $pending_i$.
		\end{enumerate}
	\end{enumerate}
	
\end{tcolorbox}

	The proof for the correctness of the algorithm is provided in Appendix \ref{app:algo}. Note that the timestamp used by the algorithm implies replica $i$ is not oblivious to update on any edge $e_{jk}\in E_i$, indicating the necessary condition in Theorem \ref{thm:Ei} is also sufficient.
	
	\textbf{Intuition for correctness:}  Our algorithm is similar to standard causal multicast algorithms \cite{birman1991lightweight}. The novelty of our algorithm lies in the {\em edge-indexed} vector timestamp, which contains a counter for each edge in the timestamp graph of the replica. Intuitively, keeping track of edges incident at $i$ ensures FIFO delivery of update messages to/from $i$, and keeping track of the other edges in $E_i$ guarantees that causal dependencies are carried when a chain of causally dependent update messages are propagated along a cycle. Although maintaining counters for all the edges in cycles for the second part is sufficient, it is not always necessary -- our $(i,e_{jk})$-loop characterizes precisely which subset of edges in the cycle is necessary and sufficient for maintaining causal consistency. 

	\textbf{Optimizations:} We briefly discuss some mechanisms to reduce the timestamp size (details in Appendix \ref{app:optimization}). 
	(1) {\em Timestamp Compression:} 
	We observe that, in our algorithm, the different elements of the vector $\tau_i$ at replica $i$ are not necessarily independent. For instance, suppose that $e_{j1},e_{j2},e_{j3},e_{j4} \in E_i$ for some $j\neq i$, with $X_{j1}=\{x\}$, $X_{j2}=\{y\}$, $X_{j3}=\{z\}$ and $X_{j4}=\{x,y,z\}$. 
    Observe that the number of updates performed to registers
    corresponding to these four edges is {\em not independent}.
    Thus, it is possible to compress the timestamp to reduce its space requirement.
	(2) {\em Allowing False Dependencies:}
	A false dependency occurs when application of an update $u_1$ is delayed at some replica, waiting for some update $u_2$ to be applied, even though $u_2\not\hb u_1$. 
	We can introduce a ``dummy'' copy of some register at replicas to change the share graph, and thus  reduce the timestamp size possibly, but at the cost of extra update messages.
	(3) {\em Restricting Inter-Replica Communication Patterns:}
	It is known that restricted communication graphs can allow dependency tracking with a lower overhead \cite{Meldal1991ExploitingLI,kulkarni2017effectiveness, bravo2017saturn} in the message-passing context. A similar observation applies in the case of partial replication too.

\subsection{Relation to previous results}\label{sec:previous}

Our results make an improvement over previous results \cite{Hlary2006AboutTE} regarding the timestamp size.
H{\'e}lary and Milani \cite{Hlary2006AboutTE} identify a larger set of edges (compared to
$E_i$ ) that replica $i$ needs to “track”, however their result, although sufficient, does not always yield the necessary set  of edges to track. Note that H{\'e}lary and Milani's results \cite{Hlary2006AboutTE} consider the client-centric causal consistency. As mentioned when introducing the replica-centric causal consistency, our necessary conditions presented in Section \ref{sec:nec} and \ref{sec:bounds} applies to their settings. 
The definition of the minimal $x$-hoop in \cite{Hlary2006AboutTE, Xphdthesis} states the following.

\begin{definition}[Hoop \cite{Hlary2006AboutTE, Xphdthesis}]
	Given a register $x$ and two replicas $r_a$ and $r_b$
	in $C(x)$ where $C(x)$ is the set of the replicas that stores $x$, we say that there is a $x$-hoop between $r_a$ and $r_b$, if there exists a path $(r_a = r_0, r_1, ... , r_k = r_b)$
	in the share graph $G$ such that:
	i) $r_h\notin C(x)$ $(1 \leq  h \leq k-1)$ and
	ii) each consecutive pair $(r_{h-1}, r_h)$ shares a register $x_h$ such that $x_h \neq x$ ($1 \leq h \leq k)$
\end{definition}
\begin{definition}[Minimal Hoop \cite{Hlary2006AboutTE, Xphdthesis}]\label{def:milani}
	
	An $x$-hoop \\ $(r_a = r_0, r_1, ... , r_k = r_b)$ is said to be minimal, if and only if 
	i) each edge of the hoop can be labelled with a different register and 
	ii) none of the edge label is shared by replica $r_a$ and $r_b$.
\end{definition}

The following result in \cite{Hlary2006AboutTE, Xphdthesis} intended to be a tight condition for achieving causal consistency.

\begin{lemma}[\cite{Hlary2006AboutTE, Xphdthesis}]\label{lemma:milani}
	A replica has to transmit some information about a register $x$
	if and only if the replica stores $x$ or belongs to a ``minimal $x$-hoop''
\end{lemma}

As we show with an example now, the condition in Lemma \ref{lemma:milani} from \cite{Hlary2006AboutTE, Xphdthesis} is not, in fact, tight \cite{test1}.

\begin{wrapfigure}{r}{0.3\textwidth}
\vspace{-\baselineskip}
	\includegraphics[width=0.3\textwidth]{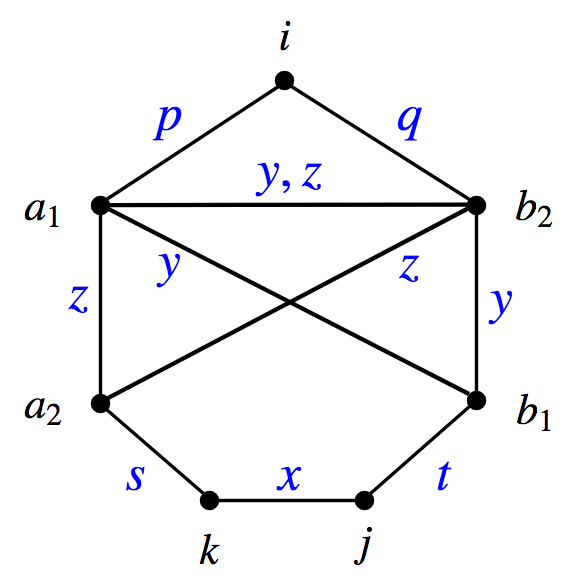}
	\caption{Example}
	\label{fig:exp}
\end{wrapfigure}

Consider the share graph in Figure \ref{fig:exp} (we omit the direction of each edge in the figure for brevity).
In the figure, the label on edges $e_{uv},e_{vu}$ shows set $X_{uv}$.
The share graph consists of replicas $i, a_1, a_2, k,j, b_1,b_2$. Replicas $j$ and $k$ share register $x$, replicas $b_1, b_2, a_1$ share register $y$, and replicas $b_2, a_1, a_2$ share register $z$.
Labels on other edges are unique and distinct from $x,y,z$.

The loop $(j, b_1, b_2,i, a_1, a_2, k)$ is
considered a ``minimal $x$-hoop'' by Definition \ref{def:milani} from \cite{Hlary2006AboutTE, Xphdthesis} because (i) the label on each
edge in the loop is distinct, (ii) none of the edge labels is shared by replica $j$ and replica $k$.
The result in H{\'e}lary and Milani \cite{Hlary2006AboutTE, Xphdthesis} implies that replica $i$
must transmit (or keep) information about updates to register $x$ by
 replicas $j,k$. However, it can be shown that presence of the two edges labeled
$y$ (and the manner they are situated) makes it unnecessary for replica
$i$ to be aware of updates to register $x$ issued by replica $j$. 
For instance, consider the execution where there is an update by replica $j$ on $x$ and then a sequence of causally dependent updates propagating along the hoop $j,b_1,b_2,i,a_1,a_2,k$ on registers $t,y,q,p,z,s$ respectively. Since the update by replica $i$ on register $p$ is also causally dependent on the update issued by replica $b_1$ on register $y$, replica $a_1$ will apply the update on $y$ before the update on $p$. Since the update on $y$ already record the dependency of the update on $x$, there is no need for replica $i$ to be aware of updates to $x$ issued by replica $j$. More details can be found in the correctness proof of the algorithm.
Similarly, replica $i$ does not need to transmit information regarding updates to $x$ issued by replica $k$.
Our necessary condition (Theorem \ref{thm:Ei}) does not require replica $i$ to keep track of these updates.

In general, our definition of the timestamp graph (Definition \ref{def:timestamp}) identifies a necessary and sufficient set of edges for each replica which is a subset of the edge set identified in \cite{Hlary2006AboutTE}.
In Appendix \ref{sec:num} later, we will numerically evaluate  the improvement compared to H{\'e}lary and Milani's results.

\section{Lower bound on Timestamp Size}\label{sec:bounds}

Section \ref{sec:nec} obtained a necessary condition on the timestamps
assigned to the replica. In this section, we obtain a lower bound (in bits)
on the size of the timestamps. In Section \ref{sec:nec}, we have the constraint that
the timestamp assigned to each replica is a function of its
causal dependency graph. From the definition of the causal dependency
graph it should be apparent that two different causal dependency graphs
may possibly correspond to the same {\em causal past} (i.e., set $S$
in Definition \ref{def:cdg}). In order to derive the lower bound, we impose the following constraint\footnote{Note that our proposed algorithm in Section \ref{sec:suf} actually satisfies this constraint.}.

\begin{constraint}\label{con:2}
	For any replica $i$,
	its timestamp at any given time is a function
	of its causal past at that time.
\end{constraint}

\begin{definition}
{\bf Timestamp space size $\sigma^i(m)$ of replica $i$ under Constraint \ref{con:2}:}
Consider the set of executions $\mathcal{E}$ in which each replica
issues up to $m$ updates.
The timestamp space size of replica $i$ under Constraint \ref{con:2}, denoted as $\sigma^i(m)$,
is the lower bound on the number of distinct timestamps that replica $i$ must assign
over all the executions in $\mathcal{E}$.
\end{definition}

Note that  replica $i$ may not use all the distinct $\sigma^i(m)$ timestamps in the same execution.
However, over all possible executions, replica $i$ will need to use at least $\sigma^i(m)$ distinct timestamps.

Let $S$ be a causal past, which is a set of updates as per Definition \ref{def:cdg}. Recall that $G=(V,E)$ denotes the share graph.
For $e_{jk}\in E$, let $S|_{e_{jk}}$ denote the set of updates in $S$ that
are issued by replica $j$ on registers in 
$X_{jk}$. For $e_{jk}\notin E$, define $S|_{e_{jk}}=\emptyset$ for convenience.

\begin{definition}[Conflict]\label{def:conf}
	Given share graph $G=(V,E)$, and two possible causal pasts $S_1,S_2$ of replica $i$ from executions in $\mathcal{E}$, 
	we say that $S_1$ and $S_2$ conflict if following conditions hold:
	\begin{enumerate}
		\setlength{\itemsep}{0pt}
		\item $\forall e\in E$, $S_1|_e\neq\emptyset\neq S_2|_e$, and
		\item  
		$\exists e\in E$ such that $S_1|_e\subset S_2|_e$, where (a) $e=e_{ij}$ or (b) $e=e_{ji}$ or (c) $\exists$ a simple loop $(i,l_1,\cdots,l_s,r_1,\cdots,r_t,i=r_{t+1})\in G$ where $e=e_{r_1 l_s}$ such that
		\begin{list}{}{}
			\item[~~(1)] $S_1|_{e_{r_p l_q}}=S_2|_{e_{r_p l_q}}$ for $1\leq p\leq t+1, 1\leq q\leq s$ and $e_{r_pl_q}\neq e_{r_1l_s}$,
			and
			\item[~~(2)] $S_x|_{e_{r_p r_{p+1}}}-\bigcup_{1\leq q\leq s}S_x|_{e_{r_p l_q}}\neq \emptyset$ 
			for $1\leq p\leq t$ and $x=1,2$
		\end{list}
	\end{enumerate}
\end{definition}

{\bf Explanation: } Condition (1) means both causal pasts $S_1,S_2$ have at least one update on every edge in the share graph, which allows us to construct executions where some replica $i$'s causal past can equal to $S_1$ or $S_2$. 
More specifically, using this property we can construct executions where the updates in $S_1$ (or $S_2$) are issued and propagated via a spanning tree rooted at replica $i$ in the share graph, thus leading to causal past $S_1$ (or $S_2$) in replica $i$.

\begin{wrapfigure}{r}{0.45\textwidth}
	\includegraphics[width=0.45\textwidth]{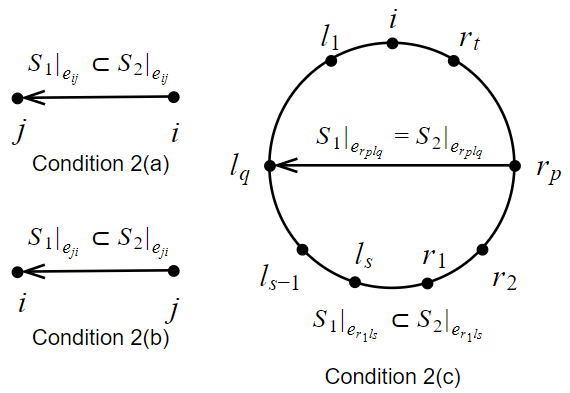}
	\caption{Condition (2) of Conflict}
	\label{fig:conflict}
\end{wrapfigure}

Condition (2) means that the set of updates in $S_1$ on some edge is a strict subset of those in $S_2$ on the same edge. As we will show in the proof, the set difference above ensures that replica must distinguish $S_1$ from $S_2$, otherwise causal consistency may be violated. Three kinds of edges are listed in the definition (see Figure \ref{fig:conflict}), (a) outgoing edges of replica $i$, (b) incoming edges of replica $i$, and (c) edges $e_{r_1l_s}$ that are in a loop which satisfies two conditions stated in the definition: (1) the set of updates on any ``chord edges'' in the loop except $e_{r_1l_s}$ are identical for $S_1$ and $S_2$, and (2) for both causal pasts $S_1$ and $S_2$, for replica $r_p$ in $r_1,...,r_t$, there exists some update sent to $r_{p+1}$ that is not sent to any of $l_1,...,l_s$. All conditions above will be used in the proof of Lemma \ref{lem:conflict}.

\begin{lemma}\label{lem:conflict}
	Consider two possible causal pasts $S_1, S_2$ of replica $i$. If $S_1$ and $S_2$  conflict, then distinct timestamps must be assigned to them for ensuring the safety and liveness properties in Definition \ref{def:causal2}.
\end{lemma}

\begin{proof}[Proof Sketch. ]
	The proof is presented in Appendix \ref{app:prlowerbound}. Here we give some intuition of the proof. First we create two executions $\mathcal{E}_1$ and $\mathcal{E}_2$, after which replica $i$ has causal past $S_1$ and $S_2$ respectively. The executions need to be created carefully such that they can be extended later to derive a contradiction as follows. If $S_1$ and $S_2$ conflict, but are assigned with the same timestamp, then replica $i$ cannot distinguish whether it has causal past $S_1$ in  $\mathcal{E}_1$ or $S_2$ in $\mathcal{E}_2$. 
	Note that $S_1$ and $S_2$ differs in updates on some edge $e$. Suppose the difference is the update set $U$ and $e=e_{jk}$. Then we can carefully extended the executions $\mathcal{E}_1$ and $\mathcal{E}_2$ such that replica $k$ with identical local timestamps $T_k$ receives an update $u$ also with identical timestamps $t_u$ in both extensions, and in one extension $u$ is causally dependent on updates in $U$ while in another it is not. Then replica $k$ cannot distinguish between the two executions, and hence may violate either safety or liveness  for causal consistency. 
\end{proof}

Once we know all the pairs of conflicting causal pasts of a replica, we can easily derive the lower bound for the timestamp space size of that replica.
For replica $i$,
we define a {\em conflict graph $H_i$} with vertex set equal to the set of all possible causal pasts of replica $i$. An edge is added between any two causal pasts of replica $i$ that conflict. 
Then, the {\em chromatic number} for the conflict graph
is a lower bound on timestamp space size.
Therefore, we have the following theorem.

\begin{theorem}\label{thm:prlowerbound}
	Consider a partially replicated shared memory system that implements replica-centric causal consistency using an algorithm under Constraint \ref{con:2}.
Let $\chi(H_i)$ denote the chromatic number of conflict graph $H_i$.
Then,  $\sigma^i(m)\geq \chi(H_i)$ for any replica $i$.
\end{theorem}

{\bf Implication:} 
Although our result does not explicitly imply a closed-form lower bound for the timestamp sizes, it can be shown that in several cases the lower bound has closed form and is tight. 
\begin{itemize}
	\setlength{\itemsep}{0pt}
	\item For instance, if the share graph is a tree, the timestamp lower bound is $2N_i\log m$ bits for replica $i$, where $N_i$ is the number of $i$'s neighbors in the share graph  and $m$ is the maximum number of updates that $i$ will issue in the execution. 
	\item  When the share graph is a cycle of $n$ replicas, the timestamp size for each replica has lower bound $2n\log m$ bits. Note that the timestamp sizes are tight in the above examples, since our algorithm will use timestamps of these sizes.
	\item In the case of {\em full replication} where the share graph is a clique and each edge shares identical set of registers, the above theorem implies the lower bound of the timestamp space size to be $m^R$ where $R$ is the total number of replicas. 
	This lower bound is also tight, because the traditional vector timestamps
	satisfy this bound
	(similar to the timestamps used by Lazy Replication \cite{Ladin1992ProvidingHA} when applied to the
	{\em peer-to-peer} architecture in Figure \ref{fig:single}).
	
\end{itemize} 
One may relate the $m^R$ lower bound for {\em full replication} to the classic lower bound on the vector clock size obtained by Charron-Bost \cite{charron1991concerning} for determining happened-before relation in a message passing system. Although two bounds equal for full replication, however, it is not true in general for partial replication. 
{\em Timestamps for deciding happened-before relation between events cannot be directly used for maintaining causal consistency and vice versa. }
One reason is that to achieve causal consistency, the timestamps should reflect information about whether there is any causally dependent update missing, but not false dependencies. Another reason is that the happened-before relation may have to be determined between any two events, while to achieve causal consistency, only for updates received by the same replica we need to determine the happened-before relation. As a result, our previous necessary and sufficient condition on the timestamps from Section \ref{sec:nec} and \ref{sec:suf} implies that the vector clock should have size equal to the number of edges in the timestamp graph (Definition \ref{def:timestamp}), which may be larger than, smaller than or equal to $n$ depending on the share graph.

\section{Extending Results to the Client-server Architecture}
\label{sec:general}

The results presented for the peer-to-peer architecture in Section \ref{sec:conditions}
can be extended to the client-server architecture.
The system model of client-server architecture is illustrated in Figure \ref{fig:multiple}.
There are $C$ clients numbered 1 through $C$. 
Each client $i$ is associated with an arbitrary subset of replicas $R_i$.
Client $i$ is restricted to perform read/write operations on registers in $\cup_{r\in R_i}X_r$.

Several natural extensions of the previous definitions are introduced in Appendix \ref{app:multiple} to obtain the results for the client-server architecture:
(a) The algorithm is extended by taking into account the fact that a client may propagate dependencies
	across two replicas. In particular, in the client-server architecture,
	a client also needs to maintain a timestamp locally, and the timestamp will be included with the request to the replicas.
(b) The share graph is augmented as shown below with additional edges that capture the causal dependencies propagation across the replicas due to the client accessing multiple replicas.
(c) The definitions of $(i,e_{jk})$-loop and timestamp graph can then be suitably
	modified to apply to the client-server architecture.
	
Below we only present the definition of augmented share graph. 
Full details of the modifications for the client-server architecture
can be found in Appendix \ref{app:multiple}.
Recall that $E$ is the set of edges in the share graph defined  previously in Section \ref{sec:pre}.
\begin{definition}[{\bf Augmented Share Graph}]
Augmented share graph $\widehat{G}$ consists of vertices in
$V=\{1,\cdots,R\}$ and directed edges in
$\widehat{E}=E\cup \{e_{jk}~|~\exists \text{ client } c \text{ such that } \,j,k \in R_c\}$.
\end{definition}
For replica $j,k$ such that $X_{jk}=\emptyset$, there is no edge
in $E$. However, if there exists client $c$ such that $j,k\in R_c$, then
directed edges between $j$ and $k$ exist in $\widehat{E}$ \footnote{The timestamps for client-server architecture will not contain the extra edges in the augmented share graph, and thus no false dependencies are introduced when we extend the results. }.

Using the augmented share graph, we can obtain a necessary condition similar to Theorem \ref{thm:Ei}, and an algorithm similar to that in Section \ref{sec:suf}, showing that the condition is also sufficient for achieving causal consistency in the client-server architecture.

\section{ Related Work}\label{sec:relatedwork}

Some of the relevant work is already discussed in
Section \ref{sec:intro}, therefore, it is not included here.

\textbf{Causal group communication:}
Several protocols \cite{birman1991lightweight, mostefaoui1993causal, kshemkalyani1998necessary} have been proposed for implementing causal group multicast with overlapping groups, and
a simulation-based evaluation on causal group multicast is presented in \cite{kalantar1999causally}. Kshemkalyani
\cite{kshemkalyani1998necessary} studied a causal group multicast protocol wherein each message $M$ is piggybacked with metadata consisting of the list of messages that happened-before $M$ and their corresponding destinations.  They investigated the necessary and sufficient conditions on the destination information tracked in this piggybacked metadata. As a result, their algorithm can remove redundant information in the metadata at run-time. However, compared to our work, their result assumes a particular structure of the metadata, and the conditions do not express how the overlapping groups (or how replicas share registers in the context of shared memory) affect the size of the metadata. 
To the best of our knowledge, lower bound for metadata size required for causality tracking with {\em overlapping} multicast groups is not previously obtained.

\textbf{Algorithms for message passing:}
The prior work on timestamps
for capturing causality in message-passing is relevant here,
in particular, several approaches for reducing timestamp size by exploiting 
communication topology information
\cite{rodrigues1995causal,Meldal1991ExploitingLI,kulkarni2017effectiveness}.
Charron-Bost proved minimum size of the vector clock is the number of the processes in the message passing system in order to capture causality \cite{charron1991concerning}.
Lower bounds on non-structured timestamps for capturing causal dependencies between events
have been studied previously \cite{giovannathesis}, but the results do not directly apply to our problem setting.
First, the events that satisfy happened-before relation in the message passing system may be false dependencies in our partial replication setting, if the event (or update) is sent to some different replica. Second, maintaining causal consistency only concerns the causality of the updates received by the same replica, not any pair of events as in message passing system in the previous works.


\textbf{Algorithms for causal consistency:}
H{\'e}lary and Milani identified the difficulty of efficient implementation under causal consistency for partial replication \cite{Hlary2006AboutTE,Xphdthesis}. 
As discussed earlier, our work improves on the results of H{\'e}lary and Milani. 
Milani has systematically studied mechanisms to implement
causal consistency, and presented a propagation based protocol for partial replication \cite{Baldoni2006OptimalPP}. Raynal \cite{raynal1998exploiting} and Birman \cite{birman1991lightweight} studied protocols for implementing partially replicated causal objects, with an architecture similar to that in Figure \ref{fig:single}, but the size of the metadata is $O(mn)$ in general, where where $n$ is the number of replicas and $m$ is the number of objects.
Shen et al. \cite{Shen2015CausalCF} proposed two algorithms,  \textsl{Full-Track} and \textsl{Opt-Track}, to achieve causal consistency for partial replication under relation $\rightarrow_{co}$ proposed by Milani \cite{Baldoni2006OptimalPP}. Their algorithms assume a particular form of the metadata, while \textsl{Full-Track} carries metadata of size $O(n^2)$ and \textsl{Opt-Track} carries metadata of optimal size.
Kshemkalyani and Hsu's work on approximate causal consistency sacrifices accuracy of causal consistencies to reduce the meta-data \cite{Kshemkalyani2015ApproximateCC, hsu2016performance}.

In a somewhat different line of research, concurrent timestamp systems for shared memory,
which enable processes to order operations using bounded timestamps have been
explored \cite{dolev1997bounded, dwork1992simple, haldar2002bounded}; the problem addressed in our work is distinct from this prior work.

\section{Summary}

This paper investigates partially replicated causally consistent shared memory systems. We present a tight necessary and sufficient condition on the replica timestamp  and a lower bound on the size of the timestamps for implementing
replica-centric causal consistency in a partially replicated system.

\bibliographystyle{abbrv}
\bibliography{ref}


\section*{Acknowledgements}
The authors thank Alessia Milani for her feedback.

	\appendix
	\appendixpage

	In the appendices, we will sometimes treat the share graph (Definition \ref{def:sg}) as an undirected graph. We will sometimes abbreviate replica-centric causal consistency as causal consistency without stating explicitly. 
	
\section{Proof of Theorem \ref{thm:Ei}}\label{sec:necproof}

	We prove Theorem \ref{thm:Ei} by showing that either safety or liveness property in Definition \ref{def:causal2} will be violated if replica $i$ is oblivious to update on any edge $e_{jk}\in E_i$.
	Consider an execution $\mathcal{E}$ in which  all issued updates have been applied at the relevant replicas, and replica $i$'s causal dependency graph is $\mathcal{R}$. This can happen since the system satisfies liveness property of the replica-centric causal consistency and all updates are applied within a finite time. 
	We will now extend the execution to show contradictions. {\em In the following extended executions, suppose any other message  that is not explicitly mentioned is delayed indefinitely.}  This is possible since the system is asynchronous.
	From Definition \ref{def:timestamp} of edge set $E_i$, there are three possible types of edges in $E_i$, which we consider in the following three cases.
	
	\begin{itemize}
		\item \textbf{Case 1: $e=e_{ij}\in E_i$:} 
		
		Let $\mathcal{E}_1$ be an extended execution where replica $i$ issues update $u_1$ on edge $e_{ij}$ (i.e., for a register in $X_{ij}$) after $\mathcal{E}$. Suppose that the causal dependency graph of $i$ after issuing $u_1$ is  $\mathcal{R}_1$.
		Let $\mathcal{E}_2$ be an extended execution where replica $i$ issues update $u_2$ on edge $e_{ij}$ after $\mathcal{E}_1$, and let $\mathcal{R}_2$ be the causal dependency graph of $i$ after issuing $u_2$.
		
		Since $\mathcal{R}_1$ and $\mathcal{R}_2$ only differ in update $u_2$ on edge $e=e_{ij}$, and replica $i$ is oblivious to update on $e_{ij}$,
		the timestamp attached to $u_1,u_2$ that
		sent to $j$ will be identical. Thus,
		replica $j$ cannot determine the correct order in which to these two updates were
		sent (recall
		that the channel is not FIFO). Thus, causal consistency cannot be assured.
		
		\item \textbf{Case 2: $e=e_{ji}\in E_i$:} 
		
		Let $\mathcal{E}_1$ be an extended execution where replica $j$ issues update $u_1$ on edge $e_{ji}$ (i.e., for a register in $X_{ij}$) after $\mathcal{E}$, but $u_1$ is not yet applied at replica $i$. Let the causal dependency graph of $i$ before applying $u_1$ be $\mathcal{R}_1$. 
		Let $\mathcal{E}_2$ be an extended execution where replica $j$ issues update $u_2$ on edge $e_{ji}$ after $\mathcal{E}_1$, and suppose that $u_1$ is applied at replica $i$ but not $u_2$. Let the new causal dependency graph of $i$ be $\mathcal{R}_2$. 
		
		Since $\mathcal{R}_1$ and $\mathcal{R}_2$ only differ in updates on edge $e=e_{ji}$, and replica $i$ is oblivious to update on $e_{ji}$, replica $i$ has identical timestamps after applying $u_1$ and before.
		Thus, when replica $i$ receives update $u_2$, it cannot differentiate
		between the following two cases: (i) $i$ has already received and applied
		update $u_1$, and thus, it can now apply update $u_2$. (ii) $i$ has not
		yet received update $u_1$, so it must wait for that update message before
		applying $u_2$. If replica $i$ applies $u_2$ when $u_2$ arrives (i.e., without
		waiting for another update message),
		but the situation is as in (ii), then safety requirement of causal consistency is violated.
		On the other hand, if replica $i$ decides to wait, but the situation is as
		in $(i)$, then another update may never be received from $j$,
		and liveness requirement of causal consistency is violated.
		
		\item \textbf{Case 3: $e=e_{jk}\in E_i$ and there exists an $(i,e_{jk})$-loop $(i,\,l_1, \cdots,l_s=k,\,j=r_1,\cdots,r_t,i)$:} 
		
		By the definition of the $(i,e_{jk})$-loop, we have
		
		\indent (i) $X_{jk}  -  \left( \cup_{1 \leq p\leq s-1} \,X_{l_p}\right)\neq\emptyset$, \\
		\indent (ii) $X_{jr_2}  -  \left( \cup_{1 \leq p\leq s-1} \,X_{l_p}\right)\neq\emptyset$, and\\
		\indent (iii) for $2\leq q\leq t$,  $X_{r_q r_{q+1}}  -  \left( \cup_{1\leq p\leq s}\, X_{l_p}\right)\neq\emptyset$.
		
		\begin{figure}[htp]
			\captionsetup[subfigure]{justification=centering}
			\centering
			\begin{subfigure}[b]{0.3\textwidth}
				\includegraphics[width=\textwidth]{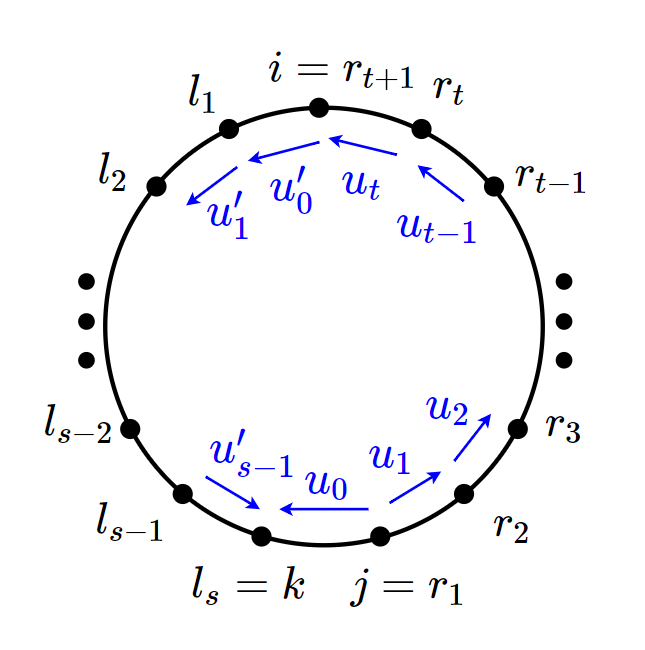}
				\caption{Illustration for Case 3.1}
				\label{fig:necex1}
			\end{subfigure}
			\begin{subfigure}[b]{0.3\textwidth}
				\includegraphics[width=\textwidth]{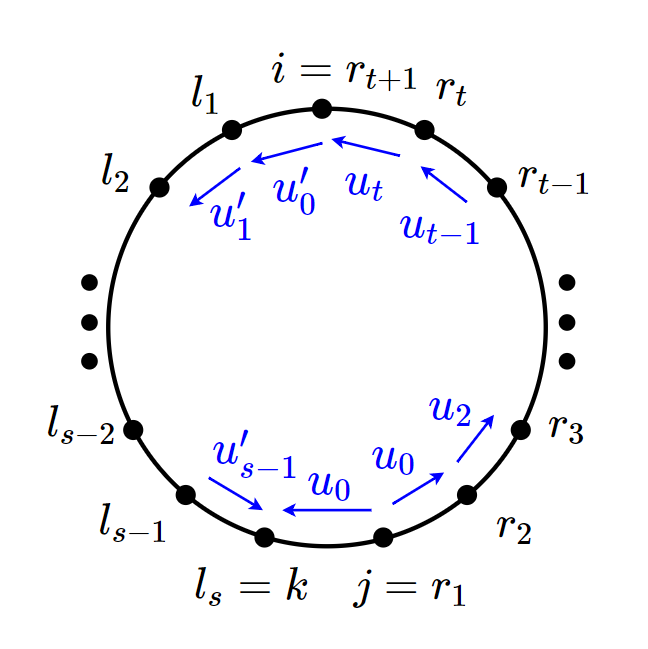}
				\caption{Illustration for Case 3.2}
				\label{fig:necex2}
			\end{subfigure}
			
			\caption{Examples of Timestamp Graphs}
			\label{fig:nec}
		\end{figure}
		
		{\bf Case 3.1:} $ X_{jr_2}  -  \left( \cup_{1 \leq p\leq s} \,X_{l_p}\right)\neq \emptyset$, that is, $X_{j r_2}$ has a register $w_1$ that is not shared by any of replicas in $l_1,\cdots, l_s$.
		
		Consider the following extension of $\mathcal{E}$ as the execution $\mathcal{E}_1$, as illustrated in Figure \ref{fig:necex1}.
		\begin{itemize}
			\item Initially, replica $r_1=j$ issues an
			update $u_0$ on edge $e_{jk}$ on register $w_0$, where $w_0\in X_{jk}  -  \left( \cup_{1 \leq p\leq s-1} \,X_{l_p}\right)$, i.e. not shared by any replicas in $l_1, \cdots, l_{s-1}$. Such $w_0$ exists since $X_{jk}  -  \left( \cup_{1 \leq p\leq s-1} \,X_{l_p}\right)\neq\emptyset$.  Thus, $u_0$ is sent to $l_s$ but not any of $l_1, \cdots, l_{s-1}$.
			
			\item 
			Replica $r_1=j$ then issues update $u_{1}$ on edge $e_{r_1 r_2}$
			on register $w_1\in X_{jr_2}  -  \left( \cup_{1 \leq p\leq s} \,X_{l_p}\right)$, i.e. $w_1$ is not shared by any replicas in $l_1, \cdots, l_s$. Thus, $u_1$ is sent to $r_2$ but not any of $l_1, \cdots, l_s$.
			The corresponding update message is next received by $r_2$.
			
			\item For $p=2$ to $t$:
			$r_p$ receives an update message from $r_{p-1}$ and applies
			the update. The update can be applied since all its causal dependencies have been applied. Then $r_p$ issues an update on edge $e_{r_p r_{p+1}}$
			on a register $w_p$ that is not shared by any of $l_1, \cdots l_s$. Such $w_p$ exists since for $2\leq q\leq t$,  $X_{r_q r_{q+1}}  -  \left( \cup_{1\leq p\leq s}\, X_{l_p}\right)\neq\emptyset$.
			Let us call this update $u_p$.
			Thus, we have constructed a sequence of updates so far such that
			$u_{0}\hb u_{1}\hb u_{2} \hb \cdots \hb u_{t}$, where $r_{t+1}=i$.
			
			\item Subsequently, $i$ issues an update $u'_{0}$ on edge $e_{i l_1}$. $l_1$ receives
			the update message, applies the update, and then issues an update $u'_{1}$
			on edge $e_{l_1 l_2}$. 
			The update can be applied since all its causal dependencies have been applied. 
			Continuing in this manner,
			we build a sequence of updates such that
			$u'_{0}\hb u'_{1}\hb\cdots \hb u'_{s-1}$, where update $u'_{p}$ in this chain is issued by
			replica $l_p$ on edge $e_{l_p l_{p+1}}$. 
			
			\item Combining the two sequences of updates, we obtain the following 
			sequence,
			$u_{0}\hb u_{1}\hb u_{2} \hb \cdots \hb u_{t} \hb
			u'_{0}\hb u'_{1}\hb u'_{2}\hb\cdots \hb u'_{s-1}$.

		\end{itemize}

		Now consider an alternate extension of $\mathcal{E}$ as the execution $\mathcal{E}_2$ in which replica $j$ does not initially
		perform update $u_{0}$, but the remaining sequence of updates above are performed.
		The timestamp of replica $i$ when issuing update $u'_{0}$ will
		be identical in both executions, because the causal dependency graphs at $i$
		when issuing update $u'_{0}$ only differ by updates on edge $e_{jk}$.

		By induction, we can easily show that the timestamp attached to the update
		$u'_{s-1}$ received by $l_s$ from $l_{s-1}$ will be identical in both
		executions. 
		{In performing the induction, we make use
			of the assumption that any other message that is not explicitly mentioned in the above executions is delayed indefinitely, including those on
			edges in $\{e_{r_x l_y} | {r_x l_y}\neq {r_1 l_s}\}$.} 
		In the first execution $u_{0}\hb u'_{s-1}$, but this is
		not the case in the second execution. If the update message from $r_1$
		to $l_s$ is not delivered before $l_s$ receives the update from $l_{s-1}$,
		then replica $l_s$ cannot determine whether it should wait for an update
		from $r_1$ or not, and either safety or liveness condition may be violated.
		
		~
		
		{\bf Case 3.2:} $ X_{jr_2}  -  \left( \cup_{1 \leq p\leq s} \,X_{l_p}\right)= \emptyset$. Since by condition (ii), $X_{jr_2}  -  \left( \cup_{1 \leq p\leq s-1} \,X_{l_p}\right)\neq\emptyset$, $\exists w_1\in X_{jr_2}\cap X_{jl_s}  -  \left( \cup_{1 \leq p\leq s-1} \,X_{l_p}\right)$, that is, $X_{j r_2}$ has a register $w_1$ that is shared by $l_s$ but not any of replicas in $l_1,\cdots, l_{s-1}$.
		
		We build two extensions of $\mathcal{E}$, similar to Case 3.1.
		Figure \ref{fig:necex2} illustrates this case.
		
		For the first execution, replica $r_1=j$ issues an update $u_{0}$ on the register $w_1$.
		Since $w_1$ is also shared by $r_2$, $u_{0}$ will be also sent  and delivered to $r_2$. Also, note that $u_0$ is sent to $k,r_2$ but not any of $l_1,\cdots,l_{s-1}$.
		Unlike Case 3.1, no other update is
		performed on edge $e_{r_1 r_2}$. The remaining sequence of updates is identical to Case 3.1. This
		results in the following happened-before relation.
		
		$u_{0}\hb u_{2} \hb u_{3} \hb \cdots \hb u_{t} \hb
		u'_{0}\hb u'_{1}\hb u'_{2}\hb\cdots \hb u'_{s-1}$
		
		For the second execution, replica $r_1$ does not issue update $u_{0}$, but the remaining sequence of updates are performed. 
		
		By similar argument as in Case 3.1, the timestamp attached to the update $u'_{s-1}$ will be identical in both executions, and replica $l_s$ cannot determine whether it should wait for an update from $r_1$ or not, and either safety or liveness condition may be violated. 
	\end{itemize}

	\section{Correctness of the Algorithm in Section \ref{sec:suf}} \label{app:algo}

	\begin{lemma}\label{lemma:single:1}
		Let $u$ be an $update(j,T,x,v)$ with timestamp $T$ from replica $j$ to replica $i$. When $\tau_i[e_{ji}]\geq T[e_{ji}]$, $u$ is already applied at replica $i$.
	\end{lemma}
	
	\begin{proof} 
		Recall Step $2(b)$ and Step $4(b)$ of the replica's algorithm in Section \ref{sec:suf}, the only way for replica $i$ to increment $\tau_i[e_{ji}]$ is by Step $4(b)$. That is,  merging $\tau_i$ with timestamp $T$ of some update, and $\tau_i[e_{ji}]$ is incremented by $1$ each time. Consider the first moment when $\tau_i[e_{ji}]=T[e_{ji}]$ after merging with $T'$ of an update $u'$. If $u'$ is issued by replica $j$, we must have $u'=u$, since $T'[e_{ji}]=T[e_{ji}]$ and both $u'$ and $u$ are on edge $e_{ji}$. Hence $u$ is applied at replica $i$. If $u'$ is issued by replica $k$ where $k\neq j$, the merge will not increase $\tau_i[e_{ji}]$, since in order to pass the predicate condition, we already have $\tau_i[e_{ji}]\geq T'[e_{ji}]$. This contradicts the assumption that it is the first moment when $\tau_i[e_{ji}]=T[e_{ji}]$. Hence $u'$ cannot be issued by replica other than $j$, which completes the proof.
		
	\end{proof}
	
	\begin{lemma}\label{lemma:single:0}
		Let $u$ be an update with timestamp $T$ from replica $j$ to $i$. Let $u'$ be an update with timestamp $T'$ from replica $k$ to $i$ such that $u'\hb u$. Then $T[e_{ki}]\geq T'[e_{ki}]$ when $k\neq j$, and $T[e_{ki}]> T'[e_{ki}]$ when $k= j$.
	\end{lemma}
	
	\begin{proof}
		When $k=j$, $u'$ and $u$ are both updating registers shared by replica $k$ and $i$. By the algorithm where write requests from client are handled, the counter of replica $k$'s timestamp on edge $e_{ki}$ is incremented by Step $2(b)$ for each write. Hence $T[e_{ki}]> T'[e_{ki}]$ when $u'\hb u$.
		
		When $k\neq j$, in order to have the happen-before relation $u'\hb u$, there exists a simple loop $(k=p_0, p_1, p_2,\cdots, p_z=j, p_{z+1}=i,k)$ where $z\geq 1$, such that each replica $p_i$ in the loop issues an update $u_i$ to the next replica $p_{i+1}$ such that $u' \hb u_0 \hb u_1 \hb \cdots \hb u_z=u$. In general, $u_0$ may equal to $u'$, but this does not affect the proof too much, and thus is omitted here for brevity. Denote the corresponding timestamps as $T',T_0,T_1,\cdots, T_z=T$ for the above updates.  Denote $e_{ki}$ as $e$ in the following context for brevity. We will prove that $T[e]\geq T'[e]$.
		
		We prove by induction on the length of the loop, where length is defined as the number of vertices in the loop. For the base case, where $z=1$ and the loop has length $3$,  consider a loop $(k,j,i)$ from $k$ to $i$. By the definition of the timestamp graph, the timestamps of replica $i,j,k$ all have a counter for edge $e$. Replica $k$ issues $u'$ to $i$, and $u_0$ to $j$. Replica $j$ issues $u$ to $i$. Since $u'\hb u_0$, $u_0$ is issued by $k$ after $u'$ is issued, and the timestamp $T_0$ for $u_0$ satisfies $T_0[e]\geq T'[e]$. Similarly, since $u_0\hb u$, $u$ is issued by $j$ after $u_0$ is applied at $j$, which by Step $4(b)$ of the algorithm we have $T[e]\geq T_0[e]$. Therefore, we have $T'[e] \leq T_0[e] \leq  T[e]$.

		Suppose for any simple loop of length $\leq h$ where $h\geq 3$, the algorithm guarantees that $T[e]\geq T'[e]$. Now consider the case where a simple loop has length $h+1$, we prove that the algorithm can guarantee $T[e]\geq T'[e]$.
		We consider the loop $\mathcal{L}=(k=p_0, p_1, p_2,\cdots, p_{h-1}=j, p_h=i, k)$ of length $h+1$, denoted as $|\mathcal{L}|=h+1$, and the chain of updates $u' \hb u_0 \hb u_1 \hb \cdots \hb u_{h-1}=u$ defined previously.
		
		\begin{figure}[h]
			\centering
			\includegraphics[width=0.3\textwidth]{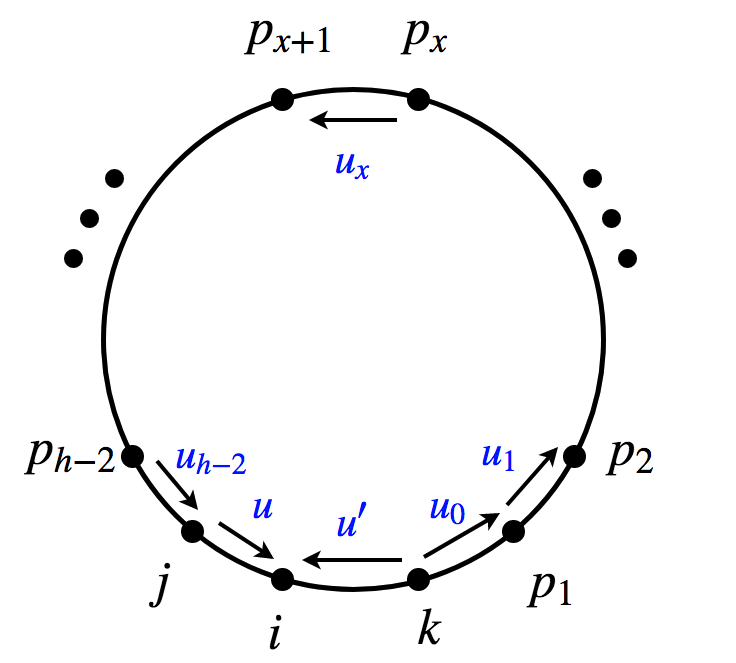}
			\caption{Illustration for loop $\mathcal{L}=(k=p_0, p_1, p_2,\cdots, p_{h-1}=j, p_h=i, k)$}
			\label{fig:algoproof}
		\end{figure}

		We follow the loop $\mathcal{L}$ starting from $k$, and see when $e$ is not contained in the timestamp graph of a replica.
		Recall that by the definition of timestamp graph, the timestamp of replica $k$ and $i$ must have a counter for $e$ because $e$ is a neighbor edge. 
		
		Suppose that the timestamp graph of $p_{l-1}$ where $1\leq l \leq h-1$ includes $e$, but the timestamp of $p_l$ does not. By the definition of timestamp graph, we can consider two cases:
		(i) $X_{ki}  -  \left( \cup_{l+1 \leq x\leq h-1} \,X_{p_x}\right)=\emptyset$, or (ii)
		there exists $p_x$ where $0\leq x\leq l-1$,  $X_{p_x p_{x+1}}  -  \left( \cup_{\substack{l+1\leq y\leq h \\ p_x p_y \neq ki}}\, X_{p_y}\right) = \emptyset$ (both case (ii) and (iii) in Definition \ref{def:ijk} are merged as case (ii) here).
		
		Consider the first case. Without loss of generality, suppose update $u'$ is an update of register $w\in X_{ki}$. Then there exists a replica $p_x$ where $l+1\leq x\leq h-1$ that shares $w$, and hence $u'$ is also on edge $e_{k p_x}$. 
		Consider the simple loop $\mathcal{L}'=(k, p_x, p_{x+1},\cdots, p_{h-1}=j, i)$ and the chain of updates $u'\hb u_x\hb\cdots \hb u_{h-1}=u$ on this loop. Since $x\geq l+1\geq 2$, we have the length of the loop $|\mathcal{L}'|\leq h$. By induction assumption on this loop, we have $T[e]\geq T'[e]$.
		
		Consider the second case. 
		Without loss of generality, suppose update $u_x$ is on register $w\in X_{p_x p_{x+1}}$. 
		Then there exists a replica $p_y$ where $l+1\leq y\leq h$ (if $x=0$, $y\leq h-1$ by definition) such that $u_{x}$ is also on edge $e_{p_x p_y}$. Then consider the simple loop $\mathcal{L}'=(k, p_1, p_2,\cdots, p_x, p_y, p_{y+1}, \cdots, i)$ and the chain of updates $u'\hb u_0\hb u_1\hb\cdots u_x \hb u_y\hb u_{y+1}\hb\cdots \hb u_{h-1}=u$ on this loop. Since $y\geq l+1\geq x+2$, we have the length of the loop $|\mathcal{L}'|\leq h$. By induction assumption on this loop, we have $T[e]\geq T'[e]$.
		
		Finally, consider the case where all replicas in the loop $\mathcal{L}=(k, p_1, p_2,\cdots, p_{h-1}=j, p_h=i, k)$ have edge $e$ in their timestamp graph. 
		Recall the chain of updates $u' \hb u_0 \hb u_1 \hb \cdots \hb u_{h-1}=u$ and their
		corresponding timestamps $T',T_0,T_1,\cdots, T_{h-1}=T$ defined previously.
		Due to the happened-before relation in $u' \hb u_0 \hb u_1 \hb \cdots \hb u_{h-1}=u$, $u_x$ is applied at replica $p_{x+1}$ before $u_{x+1}$ is issued by replica $p_{x+1}$ for $0\leq x\leq h-2$. 
		By Step $4(b)$ of our algorithm, we have $T_x[e]\leq T_{x+1}[e]$ for $0\leq x\leq h-2$, which implies $T_0[e]\leq T_{h-1}[e]=T[e]$. Since $u' \hb u_0$, $u_0$ is issued after $u'$ is issued, we have $T'[e]\leq T_0[e]$, which proves that $T'[e]\leq T[e]$.

		Therefore, for all the cases, we proved that $T'[e]\leq T[e]$ for a simple loop of length $ h+1$. 
		By induction, the algorithm can guarantee that $T[e]\geq T'[e]$ for any loop $(k, p_1, p_2,\cdots, j, i, k)$.
	\end{proof}

	\begin{lemma}\label{lemma:single:2}
		When an update $u$ is applied by replica $i$, any update $u'$ on register $x\in X_i$ such that $u'\hb u$ is already applied in the replica. 
	\end{lemma}
	
	\begin{proof}
		If $u$ is issued by another replica $j$ and propagated to replica $i$, let $u'$ be an update in the causal past of $u$. That is, $u'\hb u$ and $u'$ is on some register that is stored by replica $i$. 
		If $u'$ is issued by replica $i$, then it is already applied at replica $i$, otherwise $u'$ will not be in the causal past of $u$. Suppose $u'$ is issued by some replica $k$ for a register in $X_{ki}$, and propagated from $k$ to $i$. Let $T$ be the timestamp of update $u$, and $T'$ be the timestamp of update $u'$.
		By Lemma \ref{lemma:single:0}, $T[e_{ki}]\geq T'[e_{ki}]$ when $k\neq j$, and $T[e_{ki}]> T'[e_{ki}]$ when $k= j$.
		
		First consider the case $k\neq j$.
		When the update $u$ passes predicate $\Tau$ at replica $i$, we have $T[e_{ki}]\leq \tau_i[e_{ki}]$ where $\tau_i$ is the timestamp of replica $i$. This implies $T'[e_{ki}]\leq T[e_{ki}]\leq \tau_i[e_{ki}]$. By Lemma \ref{lemma:single:1}, $u'$ is already applied in the replica $i$.
		
		Then consider the case $k= j$.
		When the update $u$ passes predicate $\Tau$, we have $T[e_{ki}]-1\leq \tau_i[e_{ki}]$ where $\tau_i$ is the timestamp of replica $i$. This also implies $T'[e_{ki}]\leq T[e_{ki}]-1\leq \tau_i[e_{ki}]$. By Lemma \ref{lemma:single:1}, $u'$ is already applied in the replica $i$.
		
		If $u$ is issued by replica $i$, let $u'$ be an update in the causal past of $u$. Similarly, if $u'$ is issued by replica $i$, it is already applied at $i$. If $u'$ is issued by some other replica $j$, then there must exists another update $u''$ issued by some replica other than $i$ such that $u'\hb u''\hb u$ and $u''$ is applied at replica $i$. Then by the same argument of the previous case, $u'$ is already applied at replica $i$.
	\end{proof}

	\begin{theorem}\label{thm:single:0}
		The algorithm in Section \ref{sec:suf} achieves replica-centric causal consistency. 
	\end{theorem}
	
	\begin{proof}
		Safety property is implied by Lemma \ref{lemma:single:2}.
		
		Liveness property can be shown as follows. According to the algorithm, replica's timestamp is updated only when an update is applied locally. The timestamp of an update $u$ is equal to the replica's timestamp when it is issued.  And the value of each edge index in the timestamp reflects the dependent updates in that edge.  Thus when all depedencies of an update $u$ from server $k$ are applied at some replica $i$, the timestamp of server $i$ satisfies that 
		$\tau_i[e_{ki}]= T[e_{ki}]-1$ since the dependent updates from $k$ has been applied, and
		$\tau_i[e_{ji}] \geq T[e_{ji}]$ since the dependent updates from other replica $j$ has been applied for each $e_{ji}\in E_i\cap E_k,~j\neq k$. Then by Step $4$ of our algorithm $u$ will be applied at server $i$.
	
	\end{proof}

	\section{Numerical Results}\label{sec:num}

	We numerically evaluate the timestamp size of our algorithm, comparing to H{\'e}lary and Milani's results mentioned in Section \ref{sec:previous}.
	The comparison in this section demonstrates the effectiveness of our timestamps.
	
	\begin{figure}[h]
		\captionsetup[subfigure]{justification=centering}
		\centering
		\begin{subfigure}[b]{0.2\textwidth}
			\includegraphics[width=\textwidth]{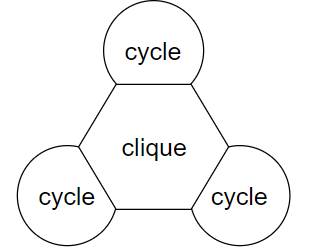}
			\caption{family of share graph}
			\label{fig:c1}
		\end{subfigure}
		\begin{subfigure}[b]{0.2\textwidth}
			\includegraphics[width=\textwidth]{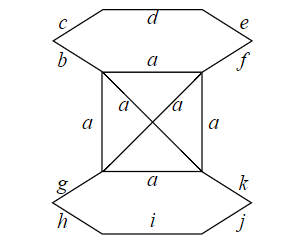}
			\caption{$x=4$, $y=6$}
			\label{fig:c2}
		\end{subfigure}
		\caption{Illustration for the share graph}
		\label{fig:clique}
	\end{figure}
	
	The family of share graph we consider is illustrated in Figure \ref{fig:c1}. The share graph consists of a clique of size $x$ ($x$ is even), and $\frac{1}{2}x$ cycles of size $y$ attached to the clique. Each cycle shares exactly one edge with the clique, and thus there are totally $x+\frac{1}{2}x\cdot(y-2)$ nodes in the graph. The labels are identical for all edges in the clique (including those shared with the cycles), and the labels are distinct for all edges in all cycles (excluding those shared with the clique). A concrete example of share graph with $x=4$ and $y=6$ is given in Figure \ref{fig:c2}.
	
	According to  our definition of the timestamp graph (Definition \ref{def:timestamp}), we can compute the timestamp size measured in the number of edge counters at each replica. Due to the existence of the clique, nodes in one cycle do not need to keep counters for directed edges in other cycles. The reason is similar to the argument for the example in Figure \ref{fig:exp}.
	For any replica in the clique (including those connected with the cycles), its timestamp graph contains all directed edges in the clique and the cycle that shares this replica. Therefore the timestamp size is 
	\[
	x(x-1)+2(y-1)
	\]
	For any replica in the cycle (excluding those connected with the clique), its timestamp graph contains all directed edges in the cycle, and incoming edges at those two replica shared by the cycle and the clique. Therefore the timestamp size is
	\[
	2(x-1)+2(y-1)
	\]
	Hence the average timestamp size of our algorithm can be computed as
	\[
	\begin{aligned}
	&\frac{x\cdot(x(x-1)+2(y-1))+(\frac{1}{2}x(y-2))\cdot(2(x-1)+2(y-1))}{x+\frac{1}{2}x(y-2)}\\
	=&2(x+y-2)+\frac{2(x^2-3x+2)}{y}
	\end{aligned}
	\]
	As for H{\'e}lary and Milani's results  \cite{Hlary2006AboutTE} (referred as original results in the following context), we can easily observe that all nodes in the graph need to keep counters for every directed edges in the share graph, thus resulting in timestamp size
	\[
	x(x-1)+2(y-1)\cdot \frac{1}{2}x=x^2-2x+xy
	\]

	\begin{figure}[htp]
		\captionsetup[subfigure]{justification=centering}
		\centering
		\begin{subfigure}[b]{0.49\textwidth}
			\includegraphics[width=\textwidth]{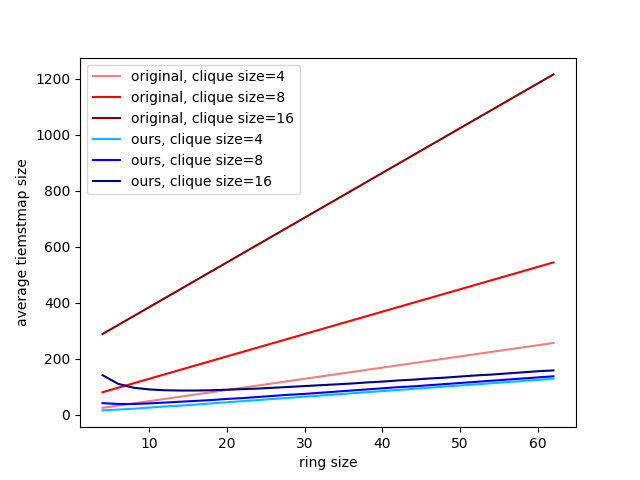}
			\caption{Comparison under different ring sizes}
			\label{fig:ringsize}
		\end{subfigure}
		\begin{subfigure}[b]{0.49\textwidth}
			\includegraphics[width=\textwidth]{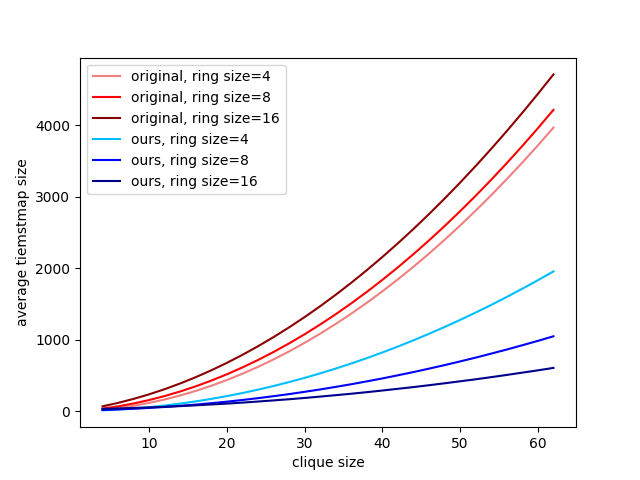}
			\caption{Comparison under different clique sizes}
			\label{fig:cliquesize}
		\end{subfigure}
		
		\caption{Comparison of average timestamp size}
		\label{fig:comparison1}
	\end{figure}

	\begin{figure}[htp]
		\captionsetup[subfigure]{justification=centering}
		\centering
		\includegraphics[width=0.5\textwidth]{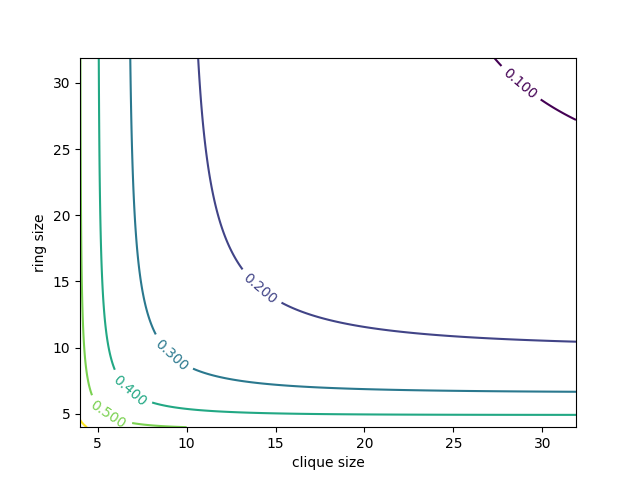}
		\caption{Timestamp size ratio}
		\label{fig:comparison2}
	\end{figure}
	We show the comparison of our timestamp size and the origin results in Figure \ref{fig:comparison1} below. In Figure \ref{fig:ringsize}, we plot how the average timestamp size changes with increasing ring sizes, while the clique size is fixed to be $4,8,16$ respectively. 
	In Figure \ref{fig:cliquesize}, we plot how the average timestamp size changes with increasing clique sizes, while the ring size is fixed to be $4,8,16$ respectively. 
	As we can observe from both figures, the original timestamp size is much larger than our timestamp size in most cases. Moreover, the original timestamp size increases significantly as either ring size or clique size grows, while the increment in our timestamp size is much smaller in comparison.
	
	To further compare the timestamp sizes of our algorithm and the original results, we compute the ratio of our timestamp size over original timestamp size, and plot the ratio with different clique sizes and ring sizes in Figure \ref{fig:comparison2}. As we can observe from the figure, the ratio reduces dramatically as ring size or clique size increases. For instance, with ring size and clique size $\geq 16$, the ratio is already less than $0.2$. We can also observe that ratio decreases fast when ring size and clique size are close to each other.

	
	\section{Proof of Lemma \ref{lem:conflict}} \label{app:prlowerbound}
	
	We will use the following terminology often: 
	\begin{itemize}
		
		
		

		\item {\em Propagating} causal past: Replica $i$ is said to propagate causal past $S$ to replica $j$ if replica $i$ send an update message to replica $j$ when the causal past of replica $i$ is $S$.

		\item {\em Update on edge $e_{ij}$ in the share graph}: An update $u$ is said to be on edge $e_{ij}$ when the update is issued by replica $i$, and the modified register is in $X_{ij}$. Thus, an update belonging to edge $e_{ij}$ will result in an update message being sent from replica $i$ to replica $j$.
		
		\item Set difference $A-B$ is defined as $A-B = \{ ~a~|~a\in A,~a\not\in B\}$.

		\item {\em Growing the causal past:} We say that, after a certain step, the causal past of a replica grows by $S$, provided
		that the causal past of the replica after that step is the union of $S$ with its casual past before the said step is performed. 
		
	\end{itemize}
	
	We also have the following observation for the proof:
	
	\textbf{Observation 1:}\label{ob1}
	To achieve causal consistency, it is necessary and sufficient that, before a replica $i$ applies an update $u_1$, it has applied any other update $u_2$ on any of its incoming edges
	such that $u_2\hb u_1$. Then the safety property of the replica-centric causal consistency is satisfied. 
	Once replica $i$ has received update messages for {\em all} the updates
	issued by $i$'s neighbors that are happened-before $u_1$, replica $i$ will eventually be able to
	apply update $u_1$. Then the liveness property of the replica-centric causal consistency is also guaranteed.
	While the order in which the updates are received by replica
	$i$ from its neighbors
	may affect how long the updates are buffered, the order does not affect the ability to apply update $u_1$
	after all the causally preceding updates are received. To reiterate,
	once all of the updates from $i$'s neighbors that happened-before $u_1$
	{are applied at $i$,}
	update $u_1$ can be applied at $i$.
	We will make use of this observation in our proofs. 
	
	Recall the statement of Definition \ref{def:conf} and Lemma \ref{lem:conflict}:
	
	\textbf{Definition \ref{def:conf}} (conflict).
	{\em
		Given share graph $G=(V,E)$, for two possible causal pasts $S_1,S_2$ of replica $i$, 
		$S_1$ and $S_2$ conflict if following conditions hold:
		\begin{enumerate}
			\item $\forall e\in E$, $S_1|_e\neq\emptyset\neq S_2|_e$, and
			\item  
			$\exists e\in E$ such that $S_1|_e\subset S_2|_e$, where $e=e_{ij}$ or $e=e_{ji}$ or $\exists$ a simple loop $(i,l_1,\cdots,l_s,r_1,\cdots,r_t,i=r_{t+1})\in G$ where $e=e_{r_1 l_s}$ such that
			\begin{list}{}{}
				\item[~~(1)] $S_1|_{e_{r_p l_q}}=S_2|_{e_{r_p l_q}}$ for $1\leq p\leq t+1, 1\leq q\leq s$ and $e_{r_pl_q}\neq e$,
				and
				\item[~~(2)] $S_x|_{e_{r_p r_{p+1}}}-\cup_{1\leq q\leq s}S_x|_{e_{r_p l_q}}\neq \emptyset$ 
				for $1\leq p\leq t$ and $x=1,2$
			\end{list}
		\end{enumerate}
	}

	\textbf{Lemma \ref{lem:conflict}}. 
	{\em
		Consider two possible causal pasts $S_1, S_2$ of replica $i$. If $S_1$ and $S_2$  conflict, then distinct timestamps must be assigned to them for ensuring the safety and liveness properties in Definition \ref{def:causal2}.
	}
	
	\paragraph*{Proof of Lemma \ref{lem:conflict}}
	
	\begin{proof}
		The proof is by contradiction. Suppose that there exists  two causal pasts of replica $i$, say $S_1$ and $S_2$, that satisfy the conditions in Lemma \ref{lem:conflict}, but both are assigned the same timestamp. 
		We will show that either safety or liveness property in Definition \ref{def:causal2} will be violated.
		By condition 1, $S_1$ and $S_2$ satisfy $|S_1|_e|\geq 1$ and $ |S_2|_e|\geq 1$ for $\forall e\in E$.
		Additionally, $S_1$ and $S_2$ satisfy condition 2 in Lemma \ref{lem:conflict}. 
		We consider each case of condition 2 separately.
		
		{\em In the following constructed executions, suppose all other message that is not explicitly mentioned is delayed indefinitely. This is possible since the system is asynchronous.} 
		\paragraph*{Case 1:} There exists $e_{ij}\in E$ such that $S_1|_{e_{ij}}\subset S_2|_{e_{ij}}$.
		
		Let $U_1=S_2|_{e_{ij}}-S_1|_{e_{ij}}$. 
		$U_1$ is non-empty because $S_1|_{e_{ij}}\subset S_2|_{e_{ij}}$.
		$U_1$ contains updates issued by replica $i$ that are on edge $e_{ij}$ in $S_2$ but not in $S_1$ (i.e., the updates
		correspond to registers in $X_i\cap X_j$).
		
		Now we construct two different executions, $\mathcal{E}_1$ and $\mathcal{E}_2$, with the following properties: After execution $\mathcal{E}_1$, replica $i$ will have causal past $S_1$, and
		after execution $\mathcal{E}_2$, replica $i$ will have causal past $S_2$. After both executions, replica $j$ will have an identical causal past, which we will name $S^*$.
		We will then extend both executions by issuing an update at replica $i$, and derive a contradiction.

		Recall from Definition \ref{def:cdg} that a causal past can be represented as a set of updates -- in particular, the happened-before relation is not
		explicitly included in the causal past.
		
		In order to create the desired executions, we will use a {\em propagation} procedure that specifies the order of
		operations performed at various replicas. This propagation procedure is presented below.
		The procedure takes as input
		a rooted spanning tree $Tree$, identifier $a$ of a replica in the
		spanning tree, and a causal past $S$ that is feasible
		at the specified replica $a$.

		In the propagation algorithm,
		$\pi(b)$ denotes the parent of $b$ in the rooted tree $Tree$. 
		
		~

		
		\begin{tcolorbox}[breakable, enhanced]

			{\bf
				Procedure $Propagation(Tree,a,S)$
			}
			
			\eIf{$a$ has at least one child in $Tree$}
			{
				\ForEach{child $c$ of $a$ in $Tree$ chosen in a predefined order}
				{
					Propagation$(Tree,c,S)$
				}
			}
			{
				Replica $a$ issues all the updates in $S|_a$ in a sequential order such that (i) the updates in $S|_a - S|_{e_{a\pi(a)}}$ are all issued
				before any update in $S|_{e_{a\pi(a)}}$ is issued, (ii) update messages sent to replicas that are {\em not ancestors} (including descendents) of $a$ in $Tree$ are not delivered until a later time (the proof will elsewhere specify when these ``held back'' update messages are delivered).
				
				\vspace*{4pt}
				
				For each ancestor $i$ of $a$, wait until all updates in $S|_{e_{ai}}$ are applied at $i$. (Note that updates in $S|_{e_{ai}}$ will be eventually applied at $i$. This is true because the dependencies of these updates are either updates issued by $a$ or by replicas in the subtree rooted at $a$ in $Tree$. Such updates have been propagated and performed at $i$ already.)
			}
			
		\end{tcolorbox}%

		\begin{figure}[htp]
			\centering
			\begin{subfigure}[b]{0.3\textwidth}
				\includegraphics[width=\textwidth]{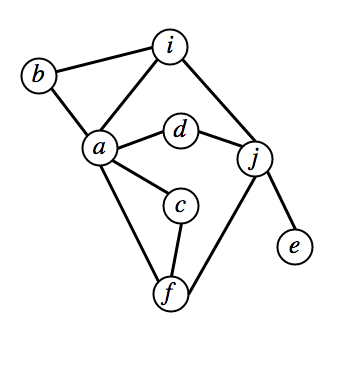}
				\caption{Share graph}
				\label{fig:sg}
			\end{subfigure}
			\begin{subfigure}[b]{0.3\textwidth}
				\includegraphics[width=\textwidth]{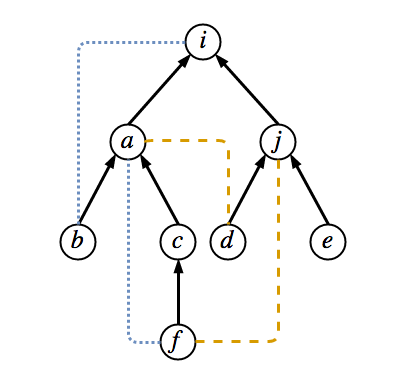}
				\caption{Spanning tree}
				\label{fig:sptree}
			\end{subfigure}
			\caption{Illustration for Propagation}\label{fig:prop}
		\end{figure}

		A simple example of the spanning tree constructed in the procedure $Propagation$ is illustrated in Figure \ref{fig:sptree}, which is based on share graph in Figure \ref{fig:sg}. The directed edges in Figure \ref{fig:sptree} represents child-father relation in the spanning tree. The blue dotted edges connect neighbors in the share graph such that one replica is an ancestor of another in the spanning tree. The brown dotted edges connect the rest of the neighbors in the share graph. The procedure $Propagation$ essentially let the replicas issue their updates in the post-order of their positions in the spanning tree, and the updates will forward along the spanning tree until reaching the root. In this example the order of replica issuing updates is $b,f,c,a,d,e,j,i$. Notice that we let the updates to ancestors to be applied, but delay those are not. For example, updates from $f$ to $a$ are applied, whereas updates from $f$ to $j$ are delayed.

		The following claim is easy to prove, due to the manner in which the updates are performed during the Propagation procedure.
		\begin{claim}
			Consider replica $b$ in the subtree of $Tree$ rooted at $a$. Then after Propagation$(Tree, a, S)$, the causal past at replica $b$ grows by $\cup_{c\in subtree_b} S|_c$ where $subtree_b$ is the subtree rooted at $b$ in $Tree$.
		\end{claim}
		The procedure $CreateExecution$ described next uses procedure $Propagation$ above. Note that the
		procedure takes edge $e_{ij}$ as input.

		
		\begin{tcolorbox}[breakable, enhanced]
			
			{\bf
				Procedure $CreateExecution(S_f,S_l,e_{ij})$
			}

			\begin{itemize}
				\item Recall that we assume that graph $G$ is connected.\footnote{Partitioned $G$ can be handled similarly
					without affecting the results.} Then there exists a spanning tree $SP$ that is rooted at replica $i$, such that
				$j$ is a child of $i$ in the spanning tree, and all the descendents of $j$ in the tree only have a path to $i$ via $j$ (namely no descendent of $j$ is a direct neighbor of $i$).

				Such a spanning tree necessarily exists because $G$ is connected, and $i$ and $j$ are neighbors in the share graph.
				The following steps are performed starting from the initial states at all the replicas.
				\begin{itemize}
					\item Replica $i$ issues updates in $S_f|_{e_{ij}}$: The corresponding update messages
					are delivered to replica $j$, and $j$ applies these updates.
					However, the update messages corresponding to these updates are not delivered to {any other replica} until
					the end of the (finite) duration of interest in this proof.

					\item  Perform procedure Propagation$(SP, i, S_f-S_f|_{e_{ij}})$ 
				\end{itemize}
				After the above steps, the causal past at $i$ is $S_f$,
				the causal past at $j$ is $S_f|_{e_{ij}} \cup \left(\cup_{b\in subtree_j}S_f|_b\right)$, and the causal past at any other replica $k$ is $\cup_{b\in subtree_k}S_f|_b$.

				\item Let $S$ be a set of updates containing at least one update on each edge of the spanning tree $SP$.

				For each child $c$ of $i$ in $SP$, perform procedure Propagation$(subtree_c, c, (S_l-S_f)\cup S)$.
				Observe that, in this step, $i$ does not issue any updates, nor apply any updates.
				
				After the above steps, the causal past at $i$ remains $S_f$, the causal past at $j$ is\\ $S_f|_{e_{ij}} \cup \left( \cup_{b\in subtree_j}(S_f\cup S_l\cup S)|_b\right) $, and that at any other replica $k$ in the spanning tree is $\cup_{b\in subtree_k}(S_f\cup S_l\cup S)|_b$.
				
				\item
				During the instantiations of the $Propagation$ procedure above,
				updates sent by neighbor $k$ of replica $j$, such that $k\neq i$
				and $k$ is not a descendent of $j$ in $SP$, are ``held back'' (i.e., delayed
				in the communication channels). 
				We now allow all of those updates to be delivered to $j$. Observation 1 ensures that these updates
				can be applied once $j$ has received all the update messages.
				After these updates have been applied, each neighbor $k\neq i$ of $j$ in $G$ ($e_{kj}$ is not necessarily an edge in the spanning tree) issues an additional update $u_k$
				on edge $e_{kj}$ that is subsequently applied at $j$. 
				$u_k$ can be applied, since all dependencies on the incoming neighbor edges of $j$ have been applied.
				These $u_k$ updates are held back
				on all other edges on which they may be propagated.
				{
					The $u_k$ updates are meant to ensure that $j$ will have in its causal
					past above all the updates from $j$'s neighbors that are also in $i$'s causal past
					$S_f$.
				}

				After this step, the causal past at $j$ is
				\begin{eqnarray*}
				&S_f|_{e_{ij}} \cup \left( \cup_{b\in subtree_j}(S_f\cup S_l\cup S)|_b\right)\cup \\
				&\left( \cup_{ \substack{e_{kj}\in G, k\neq i \\ b\in subtree_k}}(S_f\cup S_l\cup S)|_b)\right)
				\cup \left(\cup_{e_{kj}\in G,k\neq i}u_k\right)
				\label{eq:j_past}
				\end{eqnarray*}

			\end{itemize} 
			
		\end{tcolorbox}
		
		
		\paragraph*{Create executions $\mathcal{E}_1$ and $\mathcal{E}_2$}
		
		We create executions $\mathcal{E}_1$ and $\mathcal{E}_2$ such that
		at the end of these executions the causal pasts of $i$ are $S_1$ and $S_2$,
		respectively, and the causal past at $j$ is identical in both cases.
		\begin{itemize}
			\item Execution $\mathcal{E}_1$ is created by performing $CreateExecution(S_1,S_2, e_{ij})$,
			i.e., $S_f=S_1$ and $S_l=S_2$.
			At the end of execution $\mathcal{E}_1$, the causal past at $i$ is $S_1$, and
			by (\ref{eq:j_past}),
			the causal past at $j$ is \\ 
			\begin{eqnarray*}
			S^*=&S_1|_{e_{ij}} \cup \left( \cup_{b\in subtree_j}(S_1\cup S_2\cup S)|_b\right)
			\cup \\ 
			&\left( \cup_{\substack{e_{kj}\in G, k\neq i \\ b\in subtree_k}}(S_1\cup S_2\cup S)|_b)\right)
			\cup \left(\cup_{e_{kj}\in G,k\neq i}u_k\right)
			\end{eqnarray*}
			
			\item Recall that $U_1=S_2|_{e_{ij}}-S_1|_{e_{ij}}$. Execution $\mathcal{E}_2$ is created by first performing $CreateExecution(S_2-U_1,S_1, e_{ij})$, followed by $i$ issuing updates in $U_1$, but with the delivery of
			the update message corresponding to $U_1$ being delayed at $j$.

			After performing $CreateExecution(S_2-U_1,S_1, e_{ij})$, the causal past at $i$ is $S_2-U_1$,
			and by (\ref{eq:j_past}),
			the causal past at $j$ is \\ 
			\hspace*{-0.3in}
			\begin{eqnarray*}
			&(S_2-U_1)|_{e_{ij}} \cup \left( \cup_{b\in subtree_j}((S_2-U_1)\cup S_1\cup S)|_b\right)
			\cup \\
			&\left( \cup_{\substack{e_{kj}\in G, k\neq i \\ b\in subtree_k}}((S_2-U_1)\cup S_1\cup S)|_b)\right)
			\cup \left(\cup_{e_{kj}\in G,k\neq i}u_k\right)
			\end{eqnarray*}
			
			Observe that $S_2$ and $S_2-U_1$ only differ over outgoing edges at replica $i$, and
			by definition of $U_1$, we have $(S_2-U_1)|_{e_{ij}}=S_1|_{e_{ij}}$. Therefore,
			the causal past at $j$ after $CreateExecution(S_2-U_1,S_1, e_{ij})$ is also \\
			\begin{eqnarray*}
			S^*=&S_1|_{e_{ij}} \cup \left( \cup_{b\in subtree_j}(S_1\cup S_2\cup S)|_b\right)
			\cup \\
			&\left( \cup_{\substack{e_{kj}\in G, k\neq i \\ b\in subtree_k}}(S_1\cup S_2\cup S)|_b)\right)
			\cup \left(\cup_{e_{kj}\in G,k\neq i}u_k\right)
			\end{eqnarray*}
			
			This is identical to the causal past at $j$ after execution $\mathcal{E}_1$.
			In $\mathcal{E}_2$, after $CreateExecution(S_2-U_1,S_1, e_{ij})$, replica $i$ performs
			updates $U_1$ on $e_{ij}$, but the update messages are not delivered to process $j$ until
			a later time. Then, after execution $\mathcal{E}_2$, the causal past of
			$i$ will be $S_2$ and the causal past at $j$ remains same as that shown above.
		\end{itemize}
		By Constraint 1, the local timestamps of replica $j$ only depends on its causal past,
		and thus, at the end of both executions above, $j$ has the same timestamp. In other words,
		replica $j$ cannot determine whether the execution is $\mathcal{E}_1$ or $\mathcal{E}_2$.
		Also, by assumption, replica $i$ assigns the same timestamps for causal pasts $S_1$
		and $S_2$, thus, replica $i$ also has the same timestamp at the end of the two executions.

		Now we extend both the executions by replica $i$ issuing an update $u^*$ on edge $e_{ij}$.
		Update message for update $u^*$ is delivered to replica $j$ -- note that the update messages
		for updates in $U_1$ in execution $\mathcal{E}_2$ have not been delivered. This is feasible
		because the communication channel is not FIFO.
		
		\paragraph*{Deriving contradiction}
		
		When update $u^*$ is received by replica $j$ from replica $i$,
		replica $j$ must decide whether it is appropriate to apply this update.
		From replica $j$'s perspective, the two executions are indistinguishable at the time
		it receives update $u^*$.
		\begin{itemize} 
			\item
			On receipt of $u^*$, if replica $j$ assumes that
			it is in execution $\mathcal{E}_1$ but the actual execution is $\mathcal{E}_2$,
			then replica $j$ may apply $u^*$ before receiving update messages in $U_1$, which will violate the safety property of the replica-centric
			causal consistency.
			\item
			On receipt of $u^*$, if replica $j$ assumes that
			it is in execution $\mathcal{E}_2$, it will wait to receive the delayed update message (corresponding to $U_1$). However,
			if the actual execution is $\mathcal{E}_1$, then replica $j$ will wait forever for these messages (which were not sent by $i$).
			Then replica $j$ will never apply update $u^*$, even if all the dependencies of $u^*$ have been applied, which violates the liveliness
			property of the replica-centric causal consistency.
		\end{itemize}
		The above contradictions show that replica $i$ must assign different timestamps for causal pasts $S_1$ and $S_2$.

		\paragraph*{Case 2: } There exists $e_{ji}\in E$ such that $S_1|_{e_{ji}}\subset S_2|_{e_{ji}}$.
		
		Let $U_2=S_2|_{e_{ji}}-S_1|_{e_{ji}}$.
		$U_2$ is non-empty because $S_1|_{e_{ji}}\subset S_2|_{e_{ji}}$.
		
		\textbf{Create executions $\mathcal{E}_3$ and $\mathcal{E}_4$}
		
		We now define two executions similar to Case 1 above, using the spanning tree $SP$ and $subtree_j$ defined
		previously in $CreateExecution$.
		\begin{itemize}
			\item Execution $\mathcal{E}_3$:
			To construct execution $\mathcal{E}_3$, first $Propagation(SP,i,S_1)$ is performed.  After this procedure, the causal past at $i$ is $S_1$, and the causal past at $j$ is $\cup_{b\in subtree_j} S_1|_b$.
			
			Next, procedure $Propagation(subtree_j,j,(S_2-S_1)\cup S)$ is performed. Note that updates in $U_2$ are issued by $j$ but not delivered to $i$ in the above procedure. After this procedure, the causal past at $i$ remains $S_1$, and the causal past at $j$ is $\cup_{b\in subtree_j} (S_1\cup S_2 \cup S)|_b$.

			

			\item Execution $\mathcal{E}_4$:
			To construct execution $\mathcal{E}_4$, first $Propagation(SP,i,S_2)$ is performed.  After this procedure, the causal past at $i$ is $S_2$, and the causal past at $j$ is $\cup_{b\in subtree_j} S_2|_b$.
			
			Next, procedure $Propagation(subtree_j,j,(S_1-S_2)\cup S)$ is performed. After this procedure, the causal past at $i$ remains $S_2$, and the causal past at $j$ is $\cup_{b\in subtree_j} (S_1\cup S_2 \cup S)|_b$.
			
		\end{itemize}
		
		By Constraint 1, the local timestamps of replica $j$ only depends on its causal past,
		and thus, at the end of both executions above, $j$ has the same timestamp. In other words,
		replica $j$ cannot determine whether the execution is $\mathcal{E}_3$ or $\mathcal{E}_4$.
		Also, by assumption, replica $i$ assigns the same timestamps for causal pasts $S_1$
		and $S_2$, thus, replica $i$ also has the same timestamp at the end of the two executions.
		
		Now we extend both the executions by replica $j$ issuing an update $u^*$ on edge $e_{ji}$.
		Update message for update $u^*$ is delivered to replica $i$  -- note that the update messages
		for updates in $U_2$ issued by replica $j$ in execution $\mathcal{E}_3$ have not been delivered at replica $i$ yet. This is feasible
		because the communication channel is not FIFO.

		\textbf{Deriving contradiction}
		
		When update $u^*$ is received by replica $i$ from replica $j$,
		replica $i$ must decide whether it is appropriate to apply this update.
		From replica $i$'s perspective, the two executions are indistinguishable at the time
		it receives update $u^*$.
		\begin{itemize} 
			\item
			On receipt of $u^*$, if replica $i$ assumes that
			it is in execution $\mathcal{E}_4$ but the actual execution is $\mathcal{E}_3$,
			then replica $i$ may apply $u^*$ before receiving update messages for $U_2$, which violates the safety property of the replica-centric
			causal consistency.
			\item
			On receipt of $u^*$, if replica $i$ assumes that
			it is in execution $\mathcal{E}_3$, it will wait to receive the delayed update message (corresponding to $U_2$). However,
			if the actual execution is $\mathcal{E}_4$, then replica $i$ will wait forever for these messages (which have been
			previously applied by $i$ already).
			Then replica $i$ will never apply update $u^*$, even if all the dependencies of $u^*$ have been applied, which violates the liveliness
			property of the replica-centric causal consistency.
		\end{itemize}
		The above contradictions show that replica $i$ must assign different timestamps for causal pasts $S_1$ and $S_2$.

		\paragraph*{Case 3: } There exists $e=e_{r_1 l_s}\in E$ and a simple loop \\ $(i,l_1,\cdots,l_s,r_1,\cdots,r_t,r_{t+1}=i)\in G$ such that
		\begin{list}{}{}
			\item[~~(1)] $S_1|_e\subset S_2|_e$ and
			\item[~~(2)] $S_1|_{e_{r_p l_q}}=S_2|_{e_{r_p l_q}}$ for $1\leq p\leq t+1, 1\leq q\leq s$ and $e_{r_pl_q}\neq e_{r_1 l_s}$,
			and
			\item[~~(3)] $S_x|_{e_{r_p r_{p+1}}}-\cup_{1\leq q\leq s}S_x|_{e_{r_p l_q}}\neq \emptyset$ 
			for $1\leq p\leq t$ and $x=1,2$
		\end{list}
		
		\begin{figure}[h]
			\centering
			\includegraphics[width=0.3\textwidth]{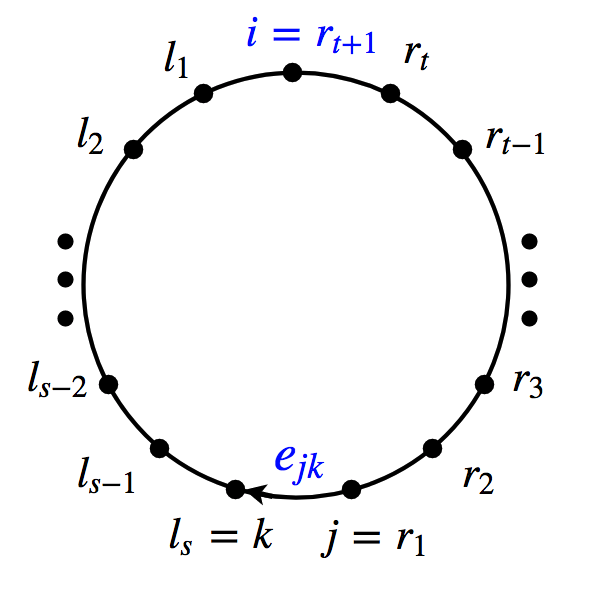}
			\caption{Illustration for Case 3}
			\label{fig:partial_ex2}
		\end{figure}
		
		Let $U_3=S_2|_e-S_1|_e$, by condition (1) above $U_3\neq \emptyset$. By condition (2) above, we can show that $U_3\cap (\cup_{1\leq q\leq s-1}S_2|_{e_{r_1 l_q}})=\emptyset$. Otherwise, if $\exists u\in U_3\cap (\cup_{1\leq q\leq s-1}S_2|_{e_{r_1 l_q}})$, then $u\in S_2|_e$ and $u\in \cup_{1\leq q\leq s-1}S_2|_{e_{r_1 l_q}}$. By condition (2), $u\in \cup_{1\leq q\leq s-1}S_1|_{e_{r_1 l_q}}$ and hence $u\in S_1|_e$, which contradicts the definition of $U_3$.
		
		The proof in this case is analogous to the proof of Case 1. Condition (2) is key to this proof.
		In particular, condition (2) makes it possible to ensure that the updates in $U_3$ appear
		in the causal past of replica $i$ without them being in the causal past replicas $l_q$, $1\leq q\leq s$. 
		We will construct two executions $\mathcal{E}_5$ and $\mathcal{E}_6$ below. These executions will satisfy the following two properties:
		\begin{itemize}
			\item The causal past of replica $i$ at the end of these executions is $S_1$ and $S_2$, respectively.
			\item The causal past of replica $l_q$, $1\leq q\leq s$, is identical after both executions, and, in particular,
			the causal past does not include $U_3$.
		\end{itemize} 
		The two executions will then be used to arrive at a contradiction, similar to Case 1.

		In graph $G$,
		there exists a directed spanning tree $T$ rooted at $i$ such that the paths $i,l_1,\cdots, l_s$,
		and  $i,r_t,\cdots, r_1$ belong to this spanning tree.
		
		Define $\chord  = \{e_{r_p l_q}~|~ 1\leq p\leq t+1, 1\leq q\leq s\}$.


		We will show how to build executions $\mathcal{E}_5$ and $\mathcal{E}_6$ below. Since the steps
		in building the two executions are quite similar, we will present the two executions together.
		\begin{enumerate}
			\item \textbf{Step 1 for $\mathcal{E}_5$ and $\mathcal{E}_6$:}
			In step 1, each replica $r_p$, {$1\leq p\leq t+1$,} issues updates in
			$S_1|_{e_{r_p l_q}}$ for $e_{r_p l_q}\in \chord$.
			From condition (2), we know that for each edge $e_{r_p l_q}\in\chord-\{e\}$,
			$S_1|_{e_{r_p l_q}} = S_2|_{e_{r_p l_q}}$.
			In both executions $\mathcal{E}_5$ and $\mathcal{E}_6$ these
			updates are applied at replicas in $l_1,l_2,\cdots, l_s$ in an identical order.
			Thus, each replica has an identical causal past at this point in both $\mathcal{E}_5$ and $\mathcal{E}_6$.
			In particular, replica $r_p$, $2\leq p\leq t+1$, has causal past
			equal to $\cup_{1\leq q\leq s}S_1|_{e_{r_p l_q}} =\cup_{1\leq q\leq s} S_2|_{e_{r_p l_q}}$.
			The causal past for $r_1$ equals $\cup_{1\leq q\leq s}S_1|_{e_{r_1 l_q}}
			= \cup_{1\leq q\leq s}S_2|_{e_{r_1 l_q}} - U_3$.

			\paragraph*
			{Step 1.1 for $\mathcal{E}_6$:}
			Replica $r_1$ issues updates in $U_3$, however, the corresponding update messages are not
			yet delivered to the recipient replicas (including $l_s$).
			By the previous argument, the updates in $U_3$ are not on any edge $e_{r_1 l_q}$ for $1\leq q\leq s$, hence the causal past for $l_1,l_2,\cdots, l_s$ remains unchanged after this step. 
			Step 1.1 only grows the causal past of $r_1$ in $\mathcal{E}_6$
			by $U_3$ to become $\cup_{1\leq q\leq s}S_2|_{e_{r_1 l_q}}$. The causal pasts of other replicas remain unchanged. \\

			\item \paragraph*{Step 2:}
			First we let all above updates from Step 1 on edges $e_{r_p r_{p+1}}$ to be delivered at $r_{p+1}$ for $1\leq p\leq t$. After that all replica still has an identical causal past, except $r_1$ and $r_2$.
			
			Recall that, by condition (3), $S_1|_{e_{r_p r_{p+1}}}-\cup_{1\leq q\leq s}S_1|_{e_{r_p l_q}}\neq \emptyset$ for $1\leq p\leq t$. Hence, in execution $\mathcal{E}_5$, there exists at least one more update in causal history $S_1$ on each edge on the path $r_1,r_2,\cdots,r_t, i$ in tree $T$ that has not been issued in the above step. Similar property holds for $S_2$ in
			execution $\mathcal{E}_6$. These properties are necessary for the desired outcome below from
			performing $Propagation$ procedure.
			(Recall that tree $T$ is defined in the earlier discussion of Case 3.)
			
			\paragraph*{Step 2 for Execution $\mathcal{E}_5$:}
			Procedure $Propagation(T,i,S_1-\cup_{\substack{e_{r_p l_q} \in \chord}} S_1|_{e_{r_p l_q}})$ is performed. After the $Propagation$ step, causal past at replica $i$ will be $S_1$. 
			
			\paragraph*{Step 2 for Execution $\mathcal{E}_6$:}
			Procedure $Propagation(T,i,S_2-\cup_{\substack{e_{r_p l_q} \in \chord}} S_2|_{e_{r_p l_q}})$ is performed. After the $Propagation$ step, causal past at replica $i$ will be $S_2$. \\

			\item \paragraph*{Step 3:} Let $S$ be a set of updates that includes at least one update
			on each edge from a child node to a parent node in the spanning tree $T$, with the constraint that
			any update on edge $e_{r_p r_{p+1}}$ occurs on a register in
			$X_{r_pr_{p+1}}-\cup_{e_{r_pl_q}\in\chord} X_{r_pl_q}$ for $1\leq p\leq t$.
			Conditions (3) ensures that such registers necessarily exist.
			The intent here is to prevent the future updates (below) at replicas $r_1,\cdots,r_t$ from affecting the causal pasts at $l_1,l_2,\cdots,l_s$.

			\paragraph*{Step 3 for Execution $\mathcal{E}_5$:}
			For each child $c$ of $i$ in tree $T$, perform procedure \\
			$Propagation(subtree_c, c, (S_2-S_1)\cup S)$,
			where $subtree_c$ is the sub-tree of $T$ rooted at $c$.
			In this step, $i$ does not issue any updates, nor perform any updates.
			
			\paragraph*{Step 3 for Execution $\mathcal{E}_6$:}
			For each child $c$ of $i$ in tree $T$, perform procedure \\
			$Propagation(subtree_c, c, (S_1-S_2)\cup S)$.
			In this step, $i$ does not issue any updates, nor perform any updates.
			
			Recall from condition (2) that $(S_1-S_2)|e_{r_p l_q}=\emptyset=(S_2-S_1)|e_{r_p l_q}$ for $r_p l_q\in\chord-\{e\}$. That is,
			{in this step}, replicas $r_p$ do not issue updates on the edges in $\chord-\{e\}$. This guarantees that information about the number of updates performed on edge $e$ {in this step} will not leak to replicas $l_1,\cdots, l_s$ in subsequent steps.

			After step 3, the causal past at $i$ remains unchanged (i.e., $S_1$ in Execution $\mathcal{E}_5$,
			and $S_2$ in $\mathcal{E}_6$).
			
			In both executions, the causal past at replica
			$d\in V-\{i,l_1,\cdots,l_s\}$ is identical, specifically,
			$$\cup_{p\in subtree_{d}} (S_1\cup S_2\cup S)|_p$$
			
			Similarly, it should be easy to see
			that the causal past at $l_q$, $1\leq q\leq s$  is also identical in both
			the executions after step 3.
			To ensure this outcome, it is important that in Step 1 of both
			executions, { the updates are issued by each $r_p$, $1\leq p \leq t+1$, in identical order.}

			\item \paragraph*{Step 4 for $\mathcal{E}_5$ and $\mathcal{E}_6$:}
			The goal in Step 4 is to ensure that the causal past of replica $l_q\in\{l_1,\cdots,l_s\}$
			includes in its causal past all the updates in $S_1\cup S_2-U_3$
			that modify the registers stored at $l_q$. 
			Except for the update messages corresponding to $U_3$ on edge $e$,
			any pending updates (from $Propagation$ procedures above) from neighbors of $l_q~(1\leq q\leq s)$ are delivered to $l_q$.

			From prior steps, observe
			that the updates in $U_3$ are only in the causal pasts of replicas $r_1,\cdots, r_t$
			in both executions, and also in the causal past of $i$ in execution $\mathcal{E}_6$.
			Note that there are no pending updates on edges in $\chord -\{e\}$, since we applied the updates on edges in $\chord-\{e\}$ in Step 1 of both executions, and no further updates on those edges are issued in step $3$.
			Thus, $U_3$ is not in the causal past of any pending updates delivered in Step 4 above.
			This, together with Observation 1 implies that these
			newly delivered updates can be applied at $l_q$ after all
			of these update messages have been delivered to $l_q$.

			The set of above updates applied at $l_q$ is identical in both executions, however,
			their causal pasts may differ in the two executions. It is because in execution $\mathcal{E}_5$ updates in $S_1$ are issued first and then those in $(S_2-S_1)\cup S$, but in execution $\mathcal{E}_6$ updates in $S_2$ are issued first and then those in $(S_1-S_2)\cup S$. The different order of how updates are issued may result in different causal past of the set of above updates delivered to $l_q$.
			To equalize the causal
			pasts at replicas $l_q$, $1\leq q\leq s$, we let each neighbor of $l_q$ except $r_1, \cdots, r_t, r_{t+1}=i$, issue one more update on the edge to $l_q$, and this update is then applied at $l_q$.
			Each above update carries the same causal past in both executions, since the causal past at the neighbors where the update is issued are identical at the time when issuing the update. Then the causal pasts at $l_q$ are equalized. 
			After these steps, causal past at replica $l_q$, $1\leq q\leq s$ is identical in both executions.
			Since, in this step, no additional update on $e_{r_p l_q}\in \chord$ is applied at $l_q$, $1\leq q\leq s$, the causal past at replicas $l_1,l_2,\cdots, l_s$ does not contain $U_3$ after both executions above.
			
			\item \paragraph*{Step 5 for $\mathcal{E}_5$ and $\mathcal{E}_6$:}
			Both executions in Step 5 issue a chain of updates $(u_0,u_1,\cdots,u_{s-1})$ along the path $(i,l_1,l_2,\cdots,l_s)$. Specifically, replica $i$ issues update $u_0$ on edge $e_{il_1}$, $l_1$ issues $u_1$ on edge $e_{l_1l_2}$, $\cdots$, $l_{s-1}$ issues update $u_{s-1}$ on edge $e_{l_{s-1} l_s}$. Observe that in $\mathcal{E}_6$, $u_{s-1}$ depends on updates in $U_3$, which means $u_{s-1}$ should be applied only after updates in $U_3$ are applied at replica $l_s$.
			
			We now argue that in both executions $\mathcal{E}_5$
			and $\mathcal{E}_6$, an identical timestamp is
			attached to the update message for $u_{s-1}$ sent by $l_{s-1}$ (i.e.,
			the timestamp for the causal past of $a_{s-1}$ when performing update $u_{s-1}$
			is identical in both executions).
			
			By Constraint 1, the timestamp of a replica only depends on its causal past. Hence the local timestamps at $l_q$ for $1\leq q\leq s$ are identical after Step 4 of $\mathcal{E}_5$ and $\mathcal{E}_6$ both.
			In step 5, when replica $i$ issues update $u_0$, the timestamp of $u_0$ is $T$ in both executions, since both causal pasts $S_1,S_2$ correspond to timestamp $T$. 
			Recall that, the dependencies of $u_0$ are already applied at $l_1$ by Step 4 of both executions, hence $u_0$ can be applied at $l_1$.
			As we mentioned previously, the timestamp at $l_1$ are identical when receiving $u_0$ in both executions, thus, the timestamp of $l_1$
			after applying $u_1$ will also be identical in both executions.
			Therefore, $l_1$ will issue the update $u_1$ with the same timestamp
			in both executions. 
			By simple induction along $l_1,l_2,\cdots, l_{s-1}$ , we know that $l_{s-1}$ will issue the update $l_{s-1}$ with the same timestamp in both executions. Recall that the causal past, and thus the local timestamp at replica $l_s$, is also identical in both executions.
			
			\textbf{Deriving contradiction}
			
			We can now derive a contradiction. On receiving $u_{s-1}$, replica $k$ cannot distinguish which execution it is in, $\mathcal{E}_5$ or $\mathcal{E}_6$, since the timestamp attached with $u_{s-1}$ and its local timestamp are identical in both executions. 
			If replica $l_s$ assumes it is in execution $\mathcal{E}_6$ (i.e., the causal past of $u_{s-1}$ contains $U_3$), it will wait for update messages for $U_3$ from replica $r_1$. However, the actual execution may be $\mathcal{E}_5$, and updates in $U_3$ may never be issued. Then replica $l_s$ will never apply $u_{s-1}$ even if the updates in the causal past of $u_{s-1}$ have already been applied, violating the liveness property of the replica-centric causal consistency.
			If replica $l_s$ assumes it is in execution $\mathcal{E}_5$ (i.e., the causal past of $u_{s-1}$ does not contain $U_3$). However, the actual execution may be $\mathcal{E}_6$, and replica $l_s$ may apply $u_{s-1}$ before receiving updates in $U_3$ from replica $r_1$, violating safety property of the replica-centric causal consistency. Hence, in both situations, replica-centric  causal consistency is violated.
		\end{enumerate}

		Therefore replica-centric causal consistency can be achieved only if $S_1$ and $S_2$ are assigned different timestamps.
	\end{proof}

	\section{Reducing the Timestamp Size in Practice} \label{app:optimization}
	
	From Section \ref{sec:nec} and \ref{sec:suf}, we know that the timestamp required to maintain causal consistency is expensive.
	In this section, we will discuss several strategies to reduce the timestamp sizes in practice. Some of these techniques exploit trade-off between timestamp size, 
	operation latency, and false dependencies.
	
	\paragraph*{Compressing timestamps:}\label{app:compress}
	We observe that the different elements of the vector $\tau_i$ at replica $i$
	are not necessarily independent.
	For instance, suppose that $e_{j1},e_{j2},e_{j3},e_{j4}$ are the only outgoing edges at $j$ that are in $E_i$, and suppose that $X_{j1}=\{x\}$, $X_{j2}=\{y\}$, $X_{j3}=\{z\}$ and $X_{j4}=\{x,y,z\}$.
	(Recall that $X_{jl}=X_j\cap X_l$.)
	The number of updates on these four edges is {\em not linearly independent}, if the numbers are {\em consistent}. Here we say numbers of updates on edges are consistent, if these numbers satisfy the linear dependency relation of the registers on edges that they corresponding to. 
	In the above example, the number of updates on edge $e_{j4}$ should be the sum of
	the number of updates
	on edges $e_{j1},e_{j2},e_{j3}$ if the numbers of updates are consistent. 
	Then, we do not need to explicitly store
	a vector element corresponding to $e_{j4}$ in the timestamp $\tau_i$. 
	In general, when the number of updates on each outgoing edges at $j$ are consistent, replica $i$ can compress its timestamp as follows.
	Let $O_j$ denote the set of outgoing edges of $j$ that are in $E_i$.
	That is, $O_j = \{ e_{jk}~|~e_{jk}\in E_i\}$.
	We identify the smallest subset of $O_j$, say $I_j$, such that
	the number of updates
	on all edges in $O_j-I_j$ can be computed as linear combinations
	of the updates on the edges in $I_j$.
	Then, for each replica $j$,
	replica $i$ only needs to store vector elements
	corresponding to the edges in $I_j$.
	To perform operations such as $merge$ and $advance$
	in the algorithm, the vector elements corresponding to edges
	in $O_j-I_j$ for each $j$ can be computed whenever needed.

	However, if the number of updates are {\em not consistent}, for instance, the number on edge $e_{j1}$ is stale while others are updated, replica $i$ cannot compress the timestamp. The above situation can happen due to the fact that the neighbor of replica $i$ may not store the counter for edge $e_{j1}$, and when it sends update to replica $i$, only the number of updates on edge $e_{j2},e_{j3},e_{j4}$ get updated at replica $i$. 
	
	More generally, for each replica $j\in V_i$, the timestamp $\tau_i$ of replica $i$ in the best case
	only needs to store $I(E_i,j)$ elements, where $I(E_i,j)$ denotes the number of maximum
	independent outgoing edges of replica $j$ that are in $E_i$. 
	When the number of updates on outgoing neighbor edges of $j$ are not consistent, replica $i$ may compress a subset of the numbers that are consistent.
	Hence the total number of elements $I'(E_i)$ in $i$'s timestamp would satisfy $I(E_i)=\sum_{j\in V_i}I(E_i,j)\leq I'(E_i)\leq |E_i|$.

	We can develop the above idea further to possibly reduce the size of each counter. Instead of counting the number of updates on all registers on each edge in set $I_j$, replica $i$ can potentially count the number of updates on only a subset of registers on that edge, thus reducing the counter size. For example, if $I_j$ contains three edges, which have registers $x$, $xy$, $xyz$ respectively, replica $i$ can simply count the number of updates on $x$, $y$ and $z$ separately, instead of counting the number of updates on $x$, $xy$ and $xyz$.

	\paragraph*{False dependencies:}
	A false dependency occurs when application of an update $u_1$ is delayed at some replica, waiting for some update $u_2$ to be applied, even though $u_2\not\hb u_1$. By allowing false dependencies to be introduced, it is possible
	to reduce timestamp size required to maintain causal consistency.
	
	Let us introduce one such approach. In our algorithm, replicas $i,j$ send updates to each other if and only if $X_{ij}\neq \emptyset$. Such
	updates contain values of updated registers in $X_{ij}$ as well as timestamps used to track causality. Now suppose that $x\in X_i$ and $x\not\in X_j$.
	Suppose that we introduce a ``dummy'' copy of register $x$ at replica $j$.
	This copy of $x$ at $j$ is ``dummy'' in the sense that no client will ever send
	a request to $j$ for an operation on $x$.
	Nevertheless, when $i$ issues an update on $x$, replica $j$ will be sent the update message, and eventually apply the update. Since $x$ is dummy at $j$, it is not really necessary to send the value (or data) associated with $x$ to $j$, and it suffices to send the timestamp (metadata) to $j$. This approach has advantages and disadvantages.
	
	\begin{itemize}
		\item To see the advantage, consider the following instantiation of the
		above approach. At each replica $j$, we introduce a dummy copy of {\em every}
		register that $j$ does not store. This effectively emulates full replication, with the important caveat that the dummy copies are never operated on.
		Thus, while the overhead of storing register copies remains identical to the
		original partial replication scheme, the timestamps can now be smaller. In particular, vector timestamps of length $R$ suffice with this emulation of full replication. 
		
		In general, instead of emulating full replication, we can use dummy register more selectively, and yet reduce the size of necessary timestamps significantly. Instead of introducing a dummy copy for every register that replica $j$ does not store, only the registers stored at $j$'s neighbors and those in the loops that pass through $j$ in the share graph are necessary. The timestamp of replica $j$ for this scheme only stores counters corresponding to neighbor replicas of $j$ and those in the loops that pass through $j$ in the shared graph. 
		As a trade-off, this solution has the following disadvantages.
		
		\item The first disadvantage is the increase in the number of update messages. In the example above, with our partial replication  algorithm, updates for register $x$ are not sent to replica $j$. However, if $j$ maintains a dummy copy of $x$, then such updates will be sent to $j$ (even if the updates contain only the metadata, or timestamps, there is still additional overhead).  
		
		The second disadvantage is the introduction of false dependencies. In the above example, suppose that replica $i$ issues update $u_1$ on $x\not\in X_{ij}$ and replica $j$ issues update $u_2$ on $y\not\in X_{ij}$. Also suppose that there are no other updates by any replica. With the original partial replication algorithm, since replicas $i,j$ will not apply each other's updates, in any execution of above
		updates, $u_1\not\hb u_2$ and $u_2\not\hb u_1$. Now suppose that $j$ maintains a dummy copy of $x$, and the update for $x$ is applied at $j$ before $j$ issues $u_2$. This will introduce the false dependency $u_1\hb u_2$. The false dependency may potentially result in additional delay in applying $u_2$ at some other replica $k$.
	\end{itemize}

	Provided the system has some guarantees on message delay, we can reduce timestamp sizes without introducing false dependencies. Consider the case where the system is {\em loosely synchronous}, which guarantees that message propagation through a path of length $\geq l$ will be slower than message propagation through one hop. In this case, replicas do not need to store counters for loops that have length $\geq l+1$, since the update travels through a long path will always arrive later than its dependent update which travels only one hop. Hence the timestamp of a replica only needs to store counters for its neighbor edges, and edges in the loops that have length $\leq l$, while guarantee that there is no false-dependency in the system. 
	
	\paragraph*{Restricting inter-replica communication patterns:} In our discussion so far, we have assumed that any pair of replicas that are adjacent in the share graph may communicate with each other directly. In the message-passing context, it is known that restricted communication graphs can allow dependency tracking with a lower overhead \cite{Meldal1991ExploitingLI,kulkarni2017effectiveness}. A similar observation applies in the case of partial replication too.
	A recent state-of-art implementation of partial replicated system \cite{bravo2017saturn} applied this idea to restrict the inter-replica communication to a shared tree, effectively reducing the size of metadata.
	
\begin{figure}
		\centering
		\includegraphics[width=0.25\textwidth]{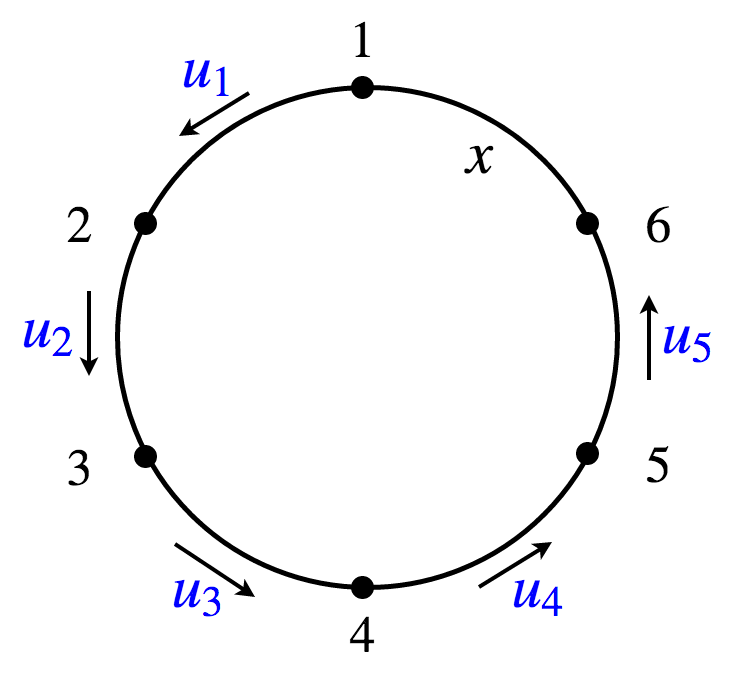}
		\caption{Illustration for ``breaking'' the ring}
		\label{fig:ring}
	\end{figure}

	We illustrate by an example how such benefits may be achieved.
	For this example, suppose that the share graph consists of a ring
	of the $R$ replica Thus,
	each replica shares a unique register with each of its neighbors
	in the ring, and does not share registers with any other replica.
	Such a ring is illustrated in Figure \ref{fig:ring} for the case of $R=6$.
	Our previous results show that if we could ``break'' the ring, the timestamp
	size may be reduced. 
	
	To achieve this goal,
	we introduce {\em virtual} registers (these have similarities to the dummy registers). The virtual registers may be shared by the different replicas in an arbitrary manner, resulting an appropriate share graph corresponding to the virtual registers.

	In our example in Figure \ref{fig:ring}, suppose that we want to break the ring by disallowing direct communication between replicas 1 and 6. However, these replicas share register $x$, thus, need to be able to send to each other updates
	to $x$. This goal can also be achieved by simulating an update message for $x$
	from replica 1 to 6, but a sequence of updates to virtual register, namely $u_i$, $i=1,2,3,4,5$ from replica $i$ to replica $i+1$, to propagate the value of $x$ from replica 1 to replica 6. When replica 6 receives update $u_5$, it would update the
	register $x$. However, with this scheme, we can redefine the share graph 
	by assuming that $x\not\in X_{16}$, while adding shared virtual registers
	between replicas $i$ and $i+1$, $1\leq i\leq 5$. In this case, we are 
	``piggybacking'' updates to $x$ on updates to virtual registers. Of course,
	since virtual registers are themselves never accessed, only metadata needs to be maintained by the virtual registers.

	In general, the assignment of the virtual registers to replicas,
	and which registers are used to piggyback updates for
	registers shared by a certain pair of replicas,  will dictate the
	communication path taken by the piggybacked update.
	
	In the extreme case, all the updates may be propagated through a single
	replica, resulting in star graph. However, more general topologies may also be created, while trading off between the overhead of the timestamps, delay in propagating updates, and false dependencies.

	\paragraph*{Sacrificing causality:}
	While the above solutions introduce false dependencies, an alternate approach that may be desirable for some applications is to sacrifice causality.
	For instance, in the timestamp graph $E_i$ defined earlier in the paper, 
	we may choose to include a smaller set of edges. In particular, for $j\neq i\neq k$, we may include edge  $e_{jk}$ only if there exists an $(i,e_{jk})$-loop
	containing at most $l+1$ edges, for some choice of $l$. Under this restriction, causal consistency will still {\em not} be violated so long as single-hop messages (or updates) are delivered faster than messages propagated over $l$ hops.
	However, when this condition does not hold, causality may be violated.

	Other approaches that sacrifice causality have been explored for full replication \cite{torres1999plausible, moore2005plausible}, and partial replication as well \cite{Kshemkalyani2015ApproximateCC, hsu2016performance}.

	\section{Results for the Client-server Architecture}\label{app:multiple}
	
	In this section, we discuss how results for the {\em peer-to-peer} architecture (Figure \ref{fig:single}) may be extended for the {\em client-server} architecture (Figure \ref{fig:multiple}).
	We extends the necessary and sufficient condition on timestamps for the peer-to-peer architecture in Section \ref{sec:conditions} to the client-server architecture. We show that with suitably modified definitions of $(i,e_{jk})$-loop and timestamp graph, we can also obtain a {\em tight condition} for the client-server architecture.

	\subsection{Replica-Centric Causal Consistency for Client-server Architecture}\label{app:causal}
	
	In the client-server architecture, clients can propagate causal dependencies of the updates when accessing different replicas. Hence we define a client-server replica-centric
	causal consistency using relation $\hb'$.

	\begin{definition}[Happened-before relation $\hb'$ for updates]
		\label{def:hb}
		Given updates $u_1$ and $u_2$, $u_1\hb' u_2$ if and only if at least one of the following conditions is true: (i) 
		$u_1$ is applied at a replica sometime before the same replica issues $u_2$.
		(ii) $u_2$ is issued by a client who previously accessed a replica that has applied $u_1$.
		(iii) There exists an update $u_3$ such that $u_1\hb' u_3$ and $u_3\hb' u_2$.
	\end{definition}
	
	Relation $\hb'$ helps us define the {\em replica-centric} causal consistency model
	of interest for the client-server architecture.

	\begin{definition}
		\label{def:causal_general}
		Replica-centric causal consistency for client-server architecture is defined using following two properties:
		
		{\bf Safety:} If an update $u_1$ for register $x\in X_i$ has been applied at a
		replica $i$, then there {\em must not exist} update $u_2$ for some register in $X_i$ such that (i)  $u_2\hb' u_1$, and (ii) replica $i$ has not yet applied $u_2$.
		
		When replica $i$ is accessed by a client, then there {\em must not exist} update $u_2$ for some register in $X_i$ such that (i) the client previously accessed a replica that has applied $u_1$, (ii) $u_2\hb' u_1$, and (iii) replica $i$ has not yet applied $u_2$.
		
		{\bf Liveness:} Any update $u$ issued by a replica $i$ for a register $x\in X_i$ will eventually be applied at each replica $j$ such that $x\in X_j$.
		
		Any write and read issued by a client to a replica will eventually return.
	\end{definition}

	\subsection{Timestamps for Replica-Centric Causal Consistency}
	Recall the definition of augmented share graph  $\widehat{G}=(\widehat{V},\widehat{E})$ defined in Section \ref{sec:general}. We can modify the definition of $(i,e_{jk})$-loop to apply for the client-server architecture as follows.
	
	\begin{definition}[Augmented $(i,e_{jk})$-loop]
		\label{def:ijk-general}
		Given replica $i$ and edge $e_{jk}$ ($j\neq i\neq k$) in augmented share graph $\widehat{G}$,
		consider a simple loop of the form
		$(i,\,l_1, \cdots,l_s=k,\,j=r_1,\cdots,r_t,i)$, where
		$s\geq 1$ and $t\geq 1$.
		Define $i=r_{t+1}$.
		The simple loop is said to be an augmented $(i,e_{jk})$-loop provided that:\\
		\indent (i) $X_{jk}  -  \left( \cup_{1 \leq p\leq s-1} \,X_{l_p}\right)\neq\emptyset$, \\
		\indent (ii)  $X_{j r_2}  -  \left( \cup_{1 \leq p\leq s-1}\, X_{l_p}\right)\neq\emptyset$ \, or \, $j,r_2\in R_c$ for some client $c$, and \\
		\indent (iii) for $2\leq q\leq t$,  $X_{r_q r_{q+1}}  -  \left( \cup_{1\leq p\leq s}\, X_{l_p}\right)\neq\emptyset$ \, or \, $r_q,r_{q+1}\in R_c$ for some client $c$,
	\end{definition}
	
	{\bf Intuition:} As mentioned before, a client communicating with different replicas propagates causal dependencies across the
	replicas. 
	This is captured in the augmented share graph by adding an edge between
	those replicas (even though the replicas may not share any registers).
	The definition of $(i,e_{jk})$-loop naturally extends to augmented share graphs, with some modification on condition (ii) and (iii). Recall the intuition of the $(i,e_{jk})$-loop is to build a dependency propagation from replica $j$ to $i$, without affecting the state of replicas $l_p$, $1\leq p\leq s-1$. When client can access multiple replicas, it can propagate the dependencies between two replicas even if they do not share any common registers. 
	
	The definition of timestamp graph can be naturally extended with the definition of augmented $(i,e_{jk})$-loop.
	
	\begin{definition}[Augmented Timestamp Graph]
		Given augmented share graph  $\widehat{G}=(\widehat{V},\widehat{E})$, augmented timestamp graph of replica $i$ is defined as a directed graph $\widehat G_i=(\widehat V_i,\widehat E_i)$, where
		\begin{eqnarray*}
			\begin{aligned}
				&\widehat E_i= (\{e_{ij}\in \widehat{E}\}\cup \{e_{ji}\in \widehat{E}\} \\ & \cup \{ e_{jk}\in \widehat{E} ~|~\exists \text{ augmented }(i,e_{jk})\text{-loop in }\widehat{G} \}) \cap E
			\end{aligned}
		\end{eqnarray*}
		\begin{eqnarray*}
		\widehat V_i=\{ u,v ~|~ e_{uv}\in E_i \}
		\end{eqnarray*}
	\end{definition}
	
	Note that $\widehat E_i$ only contains directed edges that also belong to the share graph $G=(V,E)$, that is, edges in the augmented share graph but not in the share graph are not contained in $\widehat E_i$.

	\subsection{A Necessary Condition}
	
	The necessity result of Theorem \ref{thm:Ei} extends to the client-server architecture, by replacing timestamp graph with augmented timestamp graph.

	We follow the same terminology used in Section \ref{sec:nec} to  obtain the following claim.
	\begin{theorem}\label{thm:Ei2}
		Consider a shared memory system that implements replica-centric causal consistency.
		Any replica $i$ must not be oblivious to update on any edge $e_{jk}\in \widehat E_i$ for ensuring the safety and liveness properties in Definition \ref{def:causal2}.
	\end{theorem}
	
	The proof of the above theorem basically follows the one for Theorem \ref{thm:Ei}, and is omitted here for brevity. 
	The above theorem implies that 
	it is necessary for replica $i$ to “keep track of” updates on edge $e_{jk}\in \widehat E_i$.
	
	\subsection{Sufficiency of Tracking Edges in Timestamp Graph}
	
	Now we present the algorithm for achieving causal consistency in client-server architecture. 
	The algorithm for the client-server architecture is similar to the one for the peer-to-peer architecture defined in Section \ref{sec:suf}, 
	with a key difference that each client $c$ now also
	maintains a timestamp $\mu_c$.
	
	\begin{itemize}
		\item
		Replica $i$'s timestamp: 
		Each replica $i$ maintains a vector timestamp $\tau_i$ that is indexed by the edges in $\widehat E_i$. For edge $e_{jk}\in \widehat E_i$, $\tau_i[e_{jk}]$ is an integer, initialized to $0$. 
		
		\item
		Client $c$'s  timestamp: 
		Each client $c$ maintains a vector timestamp $\mu_c$ that is indexed by the edges in $\cup_{i\in R_c} \widehat E_i$, namely all edges in the union of augmented timestamp graphs of all the replicas in $R_i$ (replicas that client $c$ can access). For edge $e_{jk}\in\cup_{i\in R_c} \widehat E_i$, $\tau_i[e_{jk}]$ is an integer, initialized to $0$. 
		
	\end{itemize}

	Recall that in the general case, each client $c$ may send its read/write request to any replica in the replica set $R_c$ associated with client $c$.
	For a replica set $R_c$, define $X_{R_c}=\cup_{i\in R_c} X_i$, that is, the set of all registers stored in replicas in $R_c$.

	\begin{tcolorbox}[breakable, enhanced]
		\textbf{Client's algorithm (for {\em client-server} architecture):} 
		
		Each client $c$ maintains a timestamp $\mu_c$, which is suitably
		initialized.
		Client $c$ may perform read/write operations on any register $x\in X_{R_c}$.
		\begin{itemize}
			\item
			When client $c$ wants to read a shared register $x\in X_{R_c}$, client $c$ sends $read(x,c,\mu_c)$ request to a replica $i\in R_c$ where $x\in X_i$; note
			that the request include client $c$'s timestamp\footnote{Instead of the actual timestamp, client $c$
				may possibly send a function of its timestamp.} and awaits replica's response containing the register value {and a timestamp $\tau$. Then client updates its timestamp using $merge_1$ function as $\mu_c=merge_1(c,\mu_c,i, \tau)$.}
			\item When client $c$ wants to write value $v$ to a shared register $x\in X_{R_c}$, client $c$ sends $write(x,v,c,\mu_c)$ request to a replica $i\in R_c$ where $x\in X_i$, and awaits the replica's response containing a timestamp $\tau$. Then client updates its timestamp  using $merge_2$ function as $\mu_c=merge_2(c,\mu_c,i, \tau)$.
			
		\end{itemize}
	\end{tcolorbox}

	\begin{tcolorbox}[breakable, enhanced]

		\textbf{Replica's algorithm (for {\em client-server} architecture):}
		
		Each replica $i$ maintains a timestamp $\tau_i$,
		which is suitably initialized.
		\begin{enumerate}
			\item When replica $i$ receives a $read(x,c,\mu)$ request from client $c$: The request is buffered until predicate $\Tau_1(i,\tau_i,c,\mu)$ evaluates true; once the predicate evaluates true, replica $i$
			responds to client $c$ with the value of the local copy of register $x$ {and its timestamp $\tau_i$}.
			
			\item When replica $i$ receives a $write(x,v,c,\mu)$ request from client $c$: The request is buffered until predicate $\Tau_2(i,\tau_i,c,\mu)$ evaluates true; once the predicate evaluates true, replica $i$ performs the following steps atomically:
			
			(i) write $v$ into the local copy of register $x$, appropriately update its timestamp $\tau_i$ using function $advance$, as $\tau_i = advance(i,\tau_i,c, \mu,x,v)$.
			
			(ii) 
			send $update(i,\tau_i,x,v)$
			to all other replicas $k$ such that $x\in X_k$, and
			
			(iii) return timestamp $\tau_i$ in the reply message to client $c$.

			\item When replica $i$ receives a message $update(k,\tau_k,x,v)$: The update is buffered until 
			predicate $\Tau_3(i,\tau_i,k,\tau_k)$ evaluates true; when the predicate evaluates true,
			replica $i$ writes value
			$v$ to its local copy of register $x$, updates its timestamp $\tau_i$ using $merge_3$ function as $\tau_i=merge_3(i,\tau_i, k, \tau_k)$, and removes $update(k,\tau_k,x,v)$ from the buffer.
			
		\end{enumerate}
		
	\end{tcolorbox}

	We now specify the predicates $\mathcal{J}_1, \mathcal{J}_2, \mathcal{J}_3$, functions $advance$ and $merge_1,merge_2,merge_3$ in the above algorithm  for the client-server architecture.

	\begin{itemize}
		\item
		Predicate
		$\Tau_1(i,\tau,c,\mu)= \Tau_2(i,\tau,c,\mu)=true$ if and only if $\tau[e_{ji}]\geq \mu[e_{ji}]$, for each $ e_{ji}\in \widehat E_i$.
		\item
		Predicate
		$\Tau_3(i, \tau,k,T)= true$ if and only if 
		
		$\tau[e_{ki}]= T[e_{ki}]-1$ ~ and ~
		$\tau[e_{ji}] \geq T[e_{ji}]$, for each $e_{ji}\in \widehat E_i \cap \widehat E_k$, $j\neq k$

		\item{ Function $advance(i,\tau,c,\mu,x,v)$} at replica $i$ returns
		vector $T_i$ (indexed by edges in $\widehat E_i$) defined as follows.
		For each $e_{jk}\in\widehat E_i$:
		
		\hspace*{0.5in}
		$T_i[e_{jk}]:= \left\{ 
		\begin{array}{l}
		\tau[e_{jk}]+1, \mbox{~if~} j=i~\mbox{and~} x\in X_{ik},\\
		\max(\tau[e_{jk}], \mu[e_{jk}]), \text{~~~~~ otherwise}
		\end{array}
		\right.
		$
		
		$advance(i,\tau,c,\mu,x,v)$ increments elements of $\tau$ corresponding to edges
		to only those replicas that also store register $x$.
		
		\item Function $merge_1(c, \mu, i, \tau)= merge_2(c, \mu, i, \tau)$ at client $c$, and returns
		following vector $T$ (indexed by edges in $\cup_{i\in R_c}\widehat E_i$):
		
		\hspace*{0.5in}
		$T[e]:= \left\{ 
		\begin{array}{l}
		\max \left(\mu[e], \tau[e]\right),
		\text{~ for each edge~} e\in \widehat E_i, \\
		\mu[e], \text{~ for each edge~} e\in (\cup_{j\in R_c}\widehat E_j)-\widehat E_i
		\end{array}
		\right.
		$
		
		\item Function $merge_3(i,T_i,k,T_k)$ at replica $i$ returns
		vector $T$ as follows:
		
		\hspace*{0.5in}
		$T[e]:= \left\{ 
		\begin{array}{l}
		\max \left(T_i[e], T_k[e]\right),
		\text{~ for each edge~} e\in \widehat E_i\cap \widehat E_k, \\
		T_i[e], \text{~ for each edge~} e\in \widehat E_i- \widehat E_k
		\end{array}
		\right.
		$
	\end{itemize}

	We omit the proof of correctness of the above generalized algorithm for brevity. 
	The algorithm and the necessary condition together imply that the augmented timestamp graph captures the {tight} edge set for any timestamp used in the algorithm for achieving causal consistency.

\end{document}